\documentclass[aps,floats]{revtex4}
\usepackage{amsmath}
\usepackage{graphicx,epsfig}

\begin{document}
\bibliographystyle {plain}

\def\oppropto{\mathop{\propto}} 
\def\opsimeq{\mathop{\simeq}}
\def\opoverderline{\mathop{\overline}}
\def\operarrow{\mathop{\longrightarrow}}
\def\opsim{\mathop{\sim}}

\def\fig#1#2{\includegraphics[height=#1]{#2}}
\def\figx#1#2{\includegraphics[width=#1]{#2}}


\title{Localization properties  
of the anomalous diffusion phase $x \sim t^{\mu}$    \\
in the directed trap model 
and in the Sinai diffusion with bias  } 

\author{C\'ecile Monthus}
\affiliation{Service de Physique Th\'eorique, 
Unit\'e de recherche associ\'ee au CNRS, \\
DSM/CEA Saclay, 91191 Gif-sur-Yvette, France}

\begin{abstract}
We study the localization properties of the anomalous diffusion
 phase $x \sim t^{\mu}$ with $0<\mu<1$ which exists both in the Sinai
 diffusion at small bias, and in the related directed trap model
 presenting a large distribution of trapping time $p(\tau) 
\sim 1/\tau^{1+\mu}$. Our starting point is the Real Space
 Renormalization method in which the whole thermal packet is 
considered to be in the same renormalized valley at large time :
 this assumption is asymptotically exact only in the limit of vanishing
 bias $\mu \to 0$ and corresponds to the Golosov localization.
For finite $\mu$, we thus generalize the usual RSRG method
 to allow for the spreading of the thermal packet over many
 renormalized valleys. Our construction allows to compute exact
 series expansions in $\mu$ for all observables : to compute observables
 at order $\mu^n$, it is sufficient to consider in each sample
a spreading of the thermal packet onto at most $(1+n)$ traps. 
So our approach provides a description of the structure of the thermal
 packet sample by sample, and a full statistical characterization of
 the important traps at a given order in $\mu$. For the directed trap model,
we show explicitly up to order $\mu^2$ how to recover 
 the exact expressions for the diffusion front, the thermal width, and
 the localization parameter $Y_2$. We then use our method to 
derive new exact results for the localization parameters $Y_k$
for arbitrary $k$, the correlation function of two particles,
 and the generating function of thermal cumulants.
We then explain how these results apply to the Sinai diffusion with bias, 
by deriving the quantitative mapping between the large-scale renormalized
descriptions of the two models. We finally study the internal structure of
the effective `traps' for the Sinai model via path-integral methods.

\end{abstract}

\maketitle



\newpage

\section{Introduction}

The motivations to study the Sinai model \cite{sinai} have two origins.
 On one hand, the Sinai model represents a `toy' disordered system,
in which many properties that exist in more complex systems
can be studied exactly, such as for instance 
aging behaviors \cite{us_sinai,dembo}, persistence exponents 
\cite{us_sinai,persistancemu}, the decoupling of the dynamics
into fast degrees of freedom which rapidly reach local equilibrium 
and a slow non-equilibrium part governed by metastable states
\cite{us_golosov}, and some chaos and rejuvenation effects \cite{sales}.
On the other hand, the Sinai model directly appears
 in various specific systems, ranging from the dynamics of domain walls
in the random field Ising chain \cite{us_rfim,corberi}
to the unzipping transition in DNA \cite{lubensky}.
It is thus interesting to obtain exact detailed informations
for various observables in the Sinai model. 

One of the most important property of the symmetric 
Sinai diffusion is the following localization phenomenon 
discovered by Golosov \cite{golosov} : 
all the thermal trajectories starting from
the same initial condition in the same sample
remain within a finite distance of each other
even in the limit of infinite time!
In particular, in a given sample, for a given initial condition,
the rescaled position $X=\frac{ x(t)}{( \ln t)^2}$
is {\it deterministic}, and it is only
after averaging over the samples that $X$ is distributed with
the Kesten distribution \cite{golosov,kesten,us_sinai}.
The physical picture is that the particle is at time $t$
near the bottom of the deepest valley it has been able to reach.
This is why the Real Space Renormalization Group method,
first introduced in the field of random quantum spin chains \cite{ma-dasgupta,daniel}, 
 is so well suited to study the 
symmetric Sinai diffusion \cite{us_sinai}.
Recently \cite{us_golosov}, we have studied
in more details the localization properties, 
by computing the infinite time limit
of the localization parameters,
which represent the 
disorder-averages of the 
probabilities that $k$ independent particles in the same
sample starting from the same initial 
condition are at the same place at time $t$
and of the correlation function $C(l,t)$,
which represents the disorder-average of the probability 
that two independent particles in the same
sample starting from the same initial condition
are at a distance $l$ from each other at time $t$.
We have moreover shown \cite{us_golosov} that the the infinite time limit
of the localization parameters and of the correlation function 
exactly coincide with the corresponding equilibrium observables 
in a Brownian potential in the thermodynamic limit.

A natural question is thus : do some of these localization properties
survive in the presence of a small bias ?

\subsection{Sinai model with bias }

The Sinai model in the presence of a constant bias $F_0>0$
can be studied in a continuum version
via the following Langevin equation \cite{annphys}
\begin{eqnarray}
\frac{dx}{dt} =   F_0 -  U'(x(t))  + \eta(t) 
\label{langevin}
\end{eqnarray}
where $\eta(t)$ is the usual thermal noise
\begin{eqnarray}
<\eta(t) \eta(t') >  = 2 T \delta(t-t') 
\end{eqnarray}
and where $U(x)$ is a Brownian random potential representing the disordered
landscape
\begin{eqnarray}
\overline {(U(x)-U(y))^2 } = 2 \sigma \delta(x-y) 
\label{defsigma}
\end{eqnarray}
Equivalently, the model may be defined by 
the Fokker-Planck equation in a given sample $\{U(x)\}$ 
for the probability distribution $P(x,t \vert x_0,0)$
\begin{eqnarray}
\partial_t P(x,t \vert x_0,0)= \partial_x \left[ T \partial_x
+U'(x)-F_0   \right] P(x,t \vert x_0,0)
\label{fokkerplanck}
\end{eqnarray}

In the biased case $F_0>0$, the diffusion becomes transient,
and there are dynamic phase transitions
\cite{kestenetal,derrida,annphys} as $F_0$ grows, in terms of
the dimensionless parameter   
\begin{eqnarray}
\mu
= \frac{F_0 T}{\sigma} 
\label{defmu}
\end{eqnarray}
For $0<\mu<1$, the mean position of the particle
presents the anomalous behavior \cite{kestenetal,derrida,annphys}
\begin{eqnarray} 
\overline{<x(t)>} \oppropto_{t \to \infty} t^{\mu}
\end{eqnarray}
whereas for $\mu>1$, there is a finite velocity $\overline{<x(t)>}
 \sim V(\mu) t$. 
For the anomalous diffusion phase, the exact diffusion front is given in
terms of L\'evy stable distributions \cite{kestenetal,annphys,tanaka,hushiyor}:
we refer the reader to the Appendix \ref{levy} for the definition and
properties of these Levy fronts.

\subsection{Directed trap model}

It has been suggested in \cite{annphys} that at large time, the physics
of the Sinai model with bias is actually equivalent to a simple directed
trap model, defined by the Master equation
\cite{aslangul}
\begin{eqnarray}
\frac{d  P_t(n)}{dt}  =  - \frac{P_t(n)}{\tau_n}+\frac{P_t(n-1)}{\tau_{n-1}}
\label{master}
\end{eqnarray}
with the initial condition $P_{t=0}(n)=\delta_{n,0}$,
and where the trapping times are independent random variables
distributed with a law presenting the algebraic decay
\begin{eqnarray} 
q(\tau) \oppropto_{\tau \to \infty} \frac{1}{ \tau^{1+\mu}} 
\label{lawtrap}
\end{eqnarray}
The anomalous diffusion phase $0<\mu<1$ then corresponds to the phase where 
the mean trapping time $<\tau> = \int d\tau \tau q(\tau)$ is infinite.
The corresponding diffusion front is also
a L\'evy diffusion front (see Appendix \ref{levy})
as for the biased Sinai diffusion discussed above.
For simplicity in the article, we choose the normalization
of the algebraic tail to be
\begin{eqnarray}
q(\tau) \opsimeq_{\tau \to \infty} \frac{\mu}{\tau^{1+\mu}} 
\label{lawrg} 
\end{eqnarray}
It is clear that this choice simply amounts to a rescaling
of $\tau$.
  
The presence of the large algebraic decay
in the effective trapping time distribution (\ref{lawtrap})
for the biased Sinai diffusion
may be understood from the Real Space Renormalization approach 
in relation with the distribution of the barriers against the drift
in the renormalized landscape at scale $\Gamma$ \cite{us_sinai}
\begin{eqnarray} 
P_{\Gamma}(F) = \theta(F-\Gamma) 2 \delta e^{- 2 \delta (F-\Gamma)} 
\label{fagainst}
\end{eqnarray}
where $2 \delta = \frac{F_0}{\sigma}$.
The trapping time $\tau \sim e^{\beta F}$ is then distributed with the power law
(\ref{lawtrap}) with the correspondence $\mu= 2 \delta T$.

\subsection{Previous results for the localization in the directed trap model}

For the directed trap model, the existing results on the extension
of the thermal packet are twofold. On one hand,  
the thermal width has been exactly computed in \cite{aslangul}
(equation (26))
\begin{eqnarray}
\overline { < \Delta n^2(t) >}  \equiv  
 \overline { \sum_{n=0}^{+\infty} n^2 P_t(n) -
[ \sum_{n=0}^{+\infty} n P_t(n) ]^2 }
= \frac{1}{\Gamma (2 \mu) } \left( \frac{ \sin \pi \mu}{\pi \mu} \right)^3 I(\mu) t^{2 \mu} 
\label{aslangul}
\end{eqnarray}
where the integral $I(\mu)$ of Equation (26) in \cite{aslangul}
can be rewritten after a change of variables as
\begin{eqnarray}
I(\mu) 
&& =  \int_0^{1} dz \frac{ (1+z) z^{\mu} (1-z)^{2 \mu} }
 {  z^{2 \mu +2} + 2 \cos \pi \mu z^{\mu+1} +1 }
\label{integralmu}
\end{eqnarray}
The result (\ref{aslangul}) shows that the the thermal
packet is spread over a length of order $t^{\mu}$.

On the other hand, the infinite-time limit
of the localization parameter
for $k=2$ has been exactly in \cite{comptejpb} :
their result (24)  
may be rewritten after a deformation of the contour
in the complex plane as
\begin{eqnarray}
Y_2 (\mu) && \equiv \lim_{t \to \infty} 
\sum_{n=0}^{+\infty} \overline{ [P_t(n)]^2 } = \int_{-\pi}^{+ \pi} \frac{ d\theta}{2 \pi}
\  \frac{e^{i \theta \mu}-e^{i \theta} }{ 1- e^{i \theta (\mu+1)}}
\label{y2exact}
\end{eqnarray}
This expression shows that $Y_2$ is finite in the full phase $0 \leq \mu<1$
and vanishes in the limit $\mu=1$. 
How can this property coexist with the result (\ref{aslangul})
for the thermal width ? The numerical
simulations of \cite{comptejpb}
show that for a single sample at fixed $t$, the probability distribution 
$P_t(n)$ is made out of a few sharp peaks that have a finite weight
but that are at a distance of order $t^{\mu}$. This explains why
at the same time, there is a finite probability to find two particles
at the same site even at infinite time, even if the thermal width 
diverges as $t^{2 \mu}$ at large time.

\subsection{Goal and results}

The aim of this article is to provide a probabilistic description,
sample by sample, 
of the localization properties of the directed trap model and of
the Sinai diffusion with bias in the anomalous diffusion phase $0<\mu<1$.
We will need to generalize the usual Real Space Renormalization group method
\cite{us_sinai} to allow for the spreading of the thermal packet
over many renormalized valleys. Indeed, in the usual RSRG method,
the whole thermal packet is considered to be in the same renormalized valley
at large time : this assumption
is asymptotically exact 
in the symmetric Sinai model and actually corresponds
to the Golosov localization \cite{golosov,us_golosov} discussed above; 
it is also valid for the biased case but only
in the double limit of vanishing bias $\mu \to 0$ and large time
with the fixed parameter $\gamma=\mu T \ln t$ \cite{us_sinai}.
We will thus define explicit rules for the
 RSRG approach with multiple valleys occupancies
and show that our construction allows to compute exact expansions
in $\mu$ for all observables. 

\subsubsection{Summary of results for the directed trap model}

For the directed trap model,
we explicitly show how to recover in a unified framework 
the expansions up to order $\mu^2$
of the exact results for the observables discussed above :

\begin{itemize}

\item  Expansion in $\mu$ of the Levy diffusion front for the rescaled variable $X=\frac{x}{t^{\mu}}$ (see Appendix \ref{levy})
\begin{eqnarray}
g(X)   
&& =  e^{-  X  }
+ \mu \gamma_{E} (X-1) e^{- X } \nonumber \\ && + \mu^2 \left[ \left(\frac{\gamma_{E}^2}{2}+ \frac{\pi^2}{12}\right)
+ X \left(\frac{\pi^2}{12} - 3 \frac{\gamma_{E}^2}{2} \right) 
+ X^2 \left(\frac{\gamma_{E}^2}{2}- \frac{\pi^2}{12}\right) \right] e^{-X}
 +O(\mu^3)
  \label{fmu}
\end{eqnarray}

\item  Expansion in $\mu$ of the thermal width 
( from (\ref{aslangul}) and (\ref{integralmu}))
\begin{eqnarray}
\Delta(\mu) \equiv \lim_{t \to \infty} \frac{\overline { < \Delta n^2 (t) >} }{ t^{2 \mu} } 
=  \mu (2 \ln 2) + \mu^2  [ - \frac{\pi^2}{6}
 + 2 \ln 2 (\ln2 -2+2 \gamma_E)  ] 
+O(\mu^3)
\label{widthexact}
\end{eqnarray}

\item Expansion in $\mu$ of the localization parameter $Y_2$ (from (\ref{y2exact}))
\begin{eqnarray}
Y_2 (\mu) = 1 - \mu (2 \ln 2) + \mu^2 ( 4 \ln 2- \frac{\pi^2}{6}) +O(\mu^3)
\label{y2exactexpansion}
\end{eqnarray}

\end{itemize}

These comparison with exact results show that our generalized RSRG procedure
is exact order by order in $\mu$ : 
to compute observables at order $\mu^n$, it is sufficient
to consider a spreading of the thermal packet onto at most $(1+n)$ traps. 
So our description provide a 
description of the structure of the thermal packet sample by sample,
and a full statistical characterization of the important traps
at a given order in $\mu$.
 
We then use our procedure to derive new exact results.
 We obtain the expansion in $\mu$ of the localization parameter $Y_k$ 
for arbitrary $k$ up to order $\mu^2$
\begin{eqnarray}
&& Y_k (\mu)  = 1+  \mu \int_0^{+\infty} \frac{dv}{v}  
 [ e^{- k v  }
+  (1-e^{-  v } )^k -1 ] \nonumber \\ && +  \mu^2 \int_0^{+\infty} \frac{dv}{v}  \int_v^{+\infty}  \frac{dw}{w} 
[ p_2^k(v,w)+p_2^k(w,v)+2 p_3^k(v,w)+1-e^{-kv} -2 (1-e^{-v})^k
 -(1-e^{-w})^k ]+O(\mu^3)
\label{ykexactexpansion}
\end{eqnarray}
where the functions $p_2$ and $p_3$ are defined in (\ref{occvw}).

 We obtain that the correlation function of two particles 
averages over the disorder reads
\begin{eqnarray}
C(l,t) && \equiv \overline{  \sum_{n=0}^{+\infty} \sum_{m=0}^{+\infty}
P(n,t\vert 0,0) P(m,t\vert 0,0) \delta_{l,\vert n-m \vert} }  \opsimeq_{t \to \infty} Y_2(\mu) \delta_{l,0}
+ \frac{1}{t^{\mu} } {\cal C}_{\mu} \left( \frac{l}{t^{\mu}} \right)
\label{correform}
 \end{eqnarray}
where the weight of the delta pic at the origin corresponds
as it should to the localization parameter $Y_2$ (\ref{y2exactexpansion}),
whereas the second part presents a scaling form of the variable $\lambda=\frac{l}{t^{\mu}}$. We obtain the following
expansion for the scaling function ${\cal C}_{\mu}$
\begin{eqnarray}
 {\cal C}_{\mu} (\lambda)= e^{-\lambda}
\left( \mu (2 \ln 2) 
+ \mu^2  \left[  \frac{\pi^2}{3}-\ln 2(4+\ln 2 +\gamma_E)
+\lambda \left( -\frac{\pi^2}{6}+\ln 2 (\ln 2 +\gamma_E) \right) \right]
 +O(\mu^3)  \right)
\label{correlong} 
\end{eqnarray}

 We also consider the generating function of 
rescaled thermal cumulants $c_k(\mu)$
\begin{eqnarray}
Z_{\mu}(s) && \equiv \overline { \ln  <e^{-s \frac{n}{t^{ \mu}}} > } 
 = \sum_{k=1}^{+\infty} \frac{(-s)^k}{k!} c_k(\mu)
\label{gencum}
 \end{eqnarray}
The first one simply represents the mean value
which can be obtained from the diffusion front (\ref{fmu})
\begin{eqnarray}
c_1(\mu) = \overline{\frac{<n>}{t^{ \mu}}} = \int_0^{+\infty} dX X f_{\mu}^{trap}(X) 
\end{eqnarray}
The second one $c_2(\mu)$ represents the thermal width $\Delta(\mu)$ (\ref{widthexact}).
We obtain the expansion at first order in $\mu$
of the generating function 
\begin{eqnarray}
 Z_{\mu}(s)  = -s +\mu \int_0^{+\infty} dY
e^{-Y} \left[ \int_0^1 \frac{dv}{v}    \ln \left[e^{- v}
+(1-e^{- v}) e^{-s Y } \right]
+ \int_1^{+\infty} \frac{dv}{v}   \ln \left[e^{- v} e^{sY}
+(1-e^{- v}) \right] \right] +O(\mu^2)
\end{eqnarray}
The series expansion in $s$
then yields all thermal cumulants at first order in $\mu$.
In particular, the first terms beyond the mean value $c_1(\mu)$
and the thermal width $c_2(\mu)$ read 
\begin{eqnarray}
c_3(\mu) && \equiv \lim_{t \to \infty} \overline{\frac{<n^3>-3<n^2><n>+2<n>^3}{t^{3 \mu}}} =   
 \mu 6 (2 \ln 3 -3 \ln 2)+O(\mu^2)
\\
c_4(\mu) && \equiv \lim_{t \to \infty} \overline{\frac{<n^4>-4<n^3><n>-3<n^2>^2+12<n^2><n>^2-6<n>^4}{t^{2 \mu}}} 
\nonumber \\
&& = \mu 24 (19 \ln 2 - 12 \ln 3)  
 +O(\mu^2)
\label{cumultrap}
\end{eqnarray}

\subsubsection{Summary of results for the biased Sinai model}

We will derive an exact quantitative mapping 
between the renormalized descriptions of the trap model
and the biased Sinai diffusion with bias.
As a consequence, in the whole anomalous diffusion phase $0<\mu<1$,
all properties of the directed trap models
that concern the rescaled quantity $X=\frac{n}{t^{\mu}}$ 
are exactly the same for the Sinai model with bias 
in terms of the rescaled quantity
\begin{eqnarray}
X=  \frac { x \sigma \beta^2 }{ \Gamma^2 (\mu) \left( t \sigma^2 \beta^3   \right)^{\mu}} 
\label{Xsinaitrap}
\end{eqnarray}
This relation was already guessed in \cite{annphys} for the special
case of the averaged diffusion fronts of the two models 
 (see Appendix \ref{levy}).
In particular, the thermal width of the Sinai model
reads from the exact result (\ref{aslangul}) of \cite{aslangul}
\begin{eqnarray}
\frac{ \overline { < \Delta x^2(t) >} }{t^{2 \mu}}
&&  = \frac {  \left(  \sigma^2 \beta^3   \right)^{2\mu}}
{  \sigma^2 \beta^4 }
\frac{\Gamma^4 (\mu)}{\Gamma (2 \mu) } \left( \frac{ \sin \pi \mu}{\pi \mu} \right)^3 I(\mu) \\
&& = \frac {  \left(  \sigma^2 \beta^3   \right)^{2\mu}}
{  \sigma^2 \beta^4 } \left[  
 \frac{(2 \ln 2)}{\mu^3} +   [ - \frac{\pi^2}{6}
 + 2 \ln 2 (\ln2 -2-2 \gamma_E)  ] \frac{1}{\mu^2}
+O(\frac{1}{\mu}) \right]
\label{widthsinai}
\end{eqnarray}
and more generally, all thermal cumulants can be obtained from
the results of the trap model (\ref{cumultrap}) via the correspondence
(\ref{Xsinaitrap}).

For the localization parameters, the result $Y_k^{trap}$ 
 represents for the biased Sinai model the probability 
to find $k$ independent particles at a finite distance of each other
in the limit of infinite time. These particles are then 
distributed with the Boltzmann distribution
in a infinitely deep biased Brownian valley, leading to
\begin{eqnarray}
Y_k^{sinai}=Y_k^{trap} Y_k^{valley}
\end{eqnarray}
where $Y_k^{valley}$, computed in (\ref{ykvalley}),
 is the localization parameter
for $k$ particles at equilibrium   
in a infinitely deep biased Brownian valley.

For the two-point correlation function, we obtain
for the biased Sinai model the two-scaling form
\begin{eqnarray}
C_{sinai}(l,t) = Y_2^{trap} C_{valley}(l)+ 
 \frac {  \sigma \beta^2 }{ \Gamma^2 (\mu) \left( t \sigma^2 \beta^3   \right)^{\mu}}
{\cal C}_{\mu} \left( \lambda= \frac { l \sigma \beta^2 }{ \Gamma^2 (\mu) \left( t \sigma^2 \beta^3   \right)^{\mu}}\right)
\label{corresinai}
\end{eqnarray}
where the first part represents the case where the
two particles are at a finite distance of each other at infinite time,
in which case their correlation $C_{valley}(l)$ is given by (\ref{correvalley}).
The second part, corresponding to the cases where the two particles
are in different renormalized valleys at infinite time, 
is exactly given by the scaling function
${\cal C}_{\mu}$ (\ref{correlong}) describing the long-range
behavior in the trap model.

\subsubsection{Organization of the paper}

We first study the directed trap model:
the Section \ref{maintrap} presents the usual RSRG
which yields all observables in the limit $\mu \to 0$;
in Section \ref{trapmu1} we explain the origin of
the spreading of the thermal packet at first order in $\mu$
and compute observables at this order ; in Section \ref{trapmu2}
we study the second order $\mu^2$; in Section \ref{hierarchie}
we explain the structure of the set of 
important traps at any given order $\mu^n$.

We then turn to the biased Sinai model:
in Section \ref{sinaibias}, we derive the quantitative mapping
between the large-scale renormalized descriptions
of the two models (the biased Sinai model and the directed trap model);
in Section \ref{internal},
we moreover characterize the internal structure of the `traps'
in the biased Sinai model by computing various statistical
properties of infinitely biased Brownian valleys.
The Section \ref{universal} contains a discussion of the
universality. Finally, the Section \ref{conclusion}
contains the conclusion, and some more technical details
are given in the Appendices.

\section{ Directed trap model in the limit $\mu \to 0$ }

\label{maintrap}

The Real Space Renormalization procedure for the Sinai model
\cite{us_sinai} can be reformulated for
the directed trap model as follows.
At time $t$, all traps with trapping time $\tau_i<t$
are decimated and replaced by a ``flat landscape''
to produce the renormalized landscape at time $t$.
We stress here that, contrary to the symmetric Sinai diffusion,
the remaining traps are just some of the initial traps, and that
their trapping time have
not been renormalized by the decimation of the small ones.
This non-renormalization of the trapping times
actually corresponds for the biased Sinai landscape
to the fact that barriers against the bias converge without rescaling
to a fixed distribution \cite{us_sinai}. 
The usual RSRG picture for the dynamics is now very simple :
 the particle starting at $t=0$ in the $n=0$ trap will be
 at time $t$ in the first trap of the renormalized landscape,
that is in the first trap having a trapping time bigger than $t$.
We will call this trap the Main Trap M.
In the usual RSRG approach, all thermal trajectories
are all in the same trap M. In particular, the probability distribution
in a given sample is a delta
\begin{eqnarray}
&& P^{(0)}_t(n) = \delta_{n,n_M} 
\label{frontM}
\end{eqnarray}
and the localization is total : there are no thermal fluctuations
\begin{eqnarray}
[\Delta n^2(t)]^{(0)}=0 
\label{thermalw0}
\end{eqnarray} 
and more generally all thermal cumulants beyond the first one vanish :
 the generating function of thermal cumulants
(\ref{gencum}) simply reads
\begin{eqnarray}
Z_{\mu}^{(0)}(s) && = -s \frac{\overline{n_M}}{t^{ \mu}} 
= - s 
\label{gen0}
 \end{eqnarray}

The two-particle correlation function is a delta
\begin{eqnarray}
 C^{(0)}(l,t) \equiv \sum_{n=0}^{+\infty} \sum_{m=0}^{+\infty}
P^{(0)}(n) P^{(0)}(m) \delta_{l,\vert n-m \vert} = \delta_{l,0}
\label{correm}
\end{eqnarray}
and the localization parameters have their maximal value
\begin{eqnarray}
 Y_k^{(0)}(t)=1 
\label{ykm}
\end{eqnarray}

The corresponding averaged diffusion front  
is thus simply given by the distribution of the position $n=n_M$ of the main trap 
\begin{eqnarray}
\overline{ P_{t}^{(0)}(n) }= \left[ 1- \int_t^{+\infty} d\tau q(\tau) \right]^{n} \int_t^{+\infty} d\tau q(\tau) 
\label{front0}
\end{eqnarray}
where the first part $[..]^{n}$ represents the probability that the first $n$ traps have a trapping time $\tau_i<t$, and where the last part
represents the probability that the $n^{\rm th}$ trap has a trapping time $\tau_i>t$. So the scaling function $g$ describing
the averaged diffusion front at large time
\begin{eqnarray}
\overline{ P_{t}(n) } 
\opsimeq_{t \to \infty}  \frac{1}{t^{\mu} } g \left( \frac{n} {t^{\mu}} \right)
\label{defgfront}
\end{eqnarray}
is given at this order by a simple exponential
\begin{eqnarray}
g^{(0)}(X) = e^{-X} 
\label{gM}
\end{eqnarray}
which indeed coincides with the limit $\mu \to 0$
 of the exact Levy front (see Appendix \ref{levy}).

So the approximation where all particles of the same thermal packet are 
in the same trap is correct only in the limit of vanishing $\mu$.
For finite $\mu$, we will have to allow for a possible dispersion of the 
thermal packet. In fact in the limit $\mu \to 0$
we have considered that the distribution of
the trapping time was infinitely broad in the following sense :
 all traps with $\tau_i<t$ were such that $\frac{\tau_i}{t} \sim 0$, 
whereas all traps with $\tau_i>t$
were such that $\frac{\tau_i}{t} \sim +\infty$. 
For finite $\mu$, we have to take into account
 that these ratios are not really zero or infinite.
We will do it order by order in $\mu$.

\begin{figure}

\centerline{\includegraphics[height=8cm]{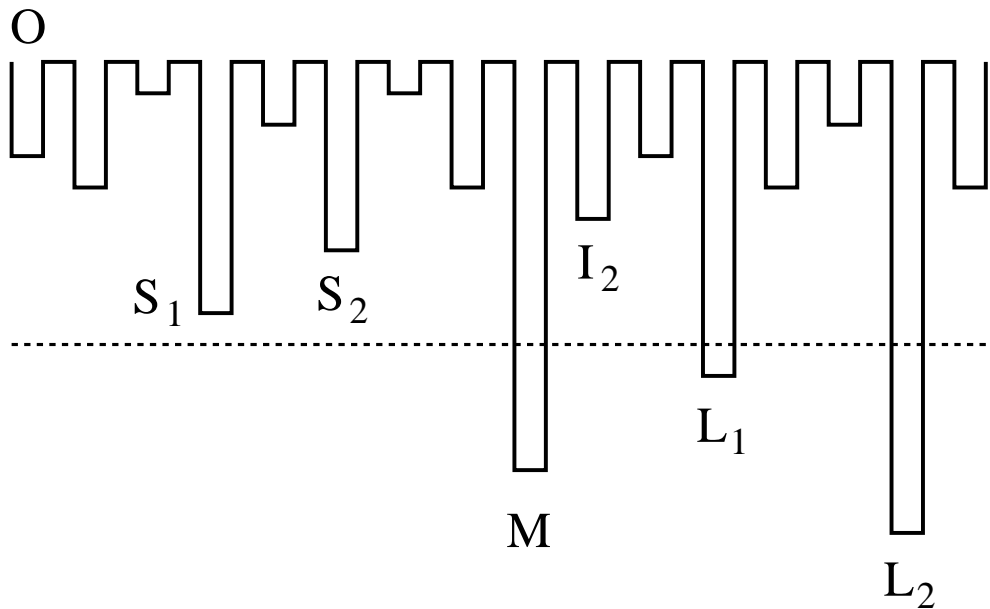}} 
\caption{ Hierarchical structure of the important traps for a particle starting at the origin.
The dashed line separates the ``small" traps (that have a trapping time
smaller than $t$) from the ``big" traps (that have a trapping time bigger 
than $t$). The first big trap called $M$ is occupied with
a weight of order $O(\mu^0)$. The next big trap $L_1$ and the biggest 
small trap $S_1$ before $M$ are occupied with weights of order $O(\mu)$.
The third big trap $L_2$, the biggest small trap $I_2$ between $M$ and $L_1$,
and the second biggest small trap $S_2$ before $M$ are occupied
with weights of order $O(\mu^2)$.  } 
\label{trap}
\end{figure}

\section{ Directed trap model at first order in $\mu$ }

\label{trapmu1}

\subsection{Origins of the dispersion of the thermal packet at order $\mu$ }

\label{dispersionmu}

At first order in $\mu$, we need to consider two effects 
(see Figure \ref{trap}) :

\begin{itemize}

\item
 the main trap M defined above has a trapping time $\tau_M$ which is not infinite. There is a small probability $(1-e^{-\frac{t}{\tau_M}})$ that the particle has already escaped from this main trap $M$
at time $t$, to jump into the next renormalized trap that we will call $L_1$
(for Large trap number 1), which is defined as the second trap satisfying $\tau_i>t$.

\item
 the biggest trap before the main trap,
that we will call $S_1$ (for Small trap number 1), 
has a trapping time $\tau_{S_1}<t$ which is not zero and thus there is a small probability $e^{-\frac{t}{\tau_{S_1}}}$ that the particle
 is still trapped in $S_1$ at time $t$.

\end{itemize}

We now describe the statistical properties of these two effects.

\subsection{ Statistical properties of the trap $L_1$}

\label{secl1}

The joint distribution of the trapping time $\tau_M$  
and of the position $n=n_M$ of the main trap $M$
and of the position $n_{L}$ of the next renormalized trap $L_1$ reads
\begin{eqnarray}
{\cal D}_{M,L_1}(n,n_{L};\tau_M) && = \theta(t<\tau_M) \theta(n<n_L)
\left[ 1- \int_{t}^{+\infty} d\tau q(\tau) \right]^{n} q(\tau_M)
\left[ 1- \int_{t}^{+\infty} d\tau q(\tau) \right]^{n_L-n-1}
 \int_t^{+\infty} d\tau q(\tau) \\
&& \opsimeq_{t \to \infty} \frac{1}{t^{2 \mu} }
D_{M,L_1} \left(X=\frac{n}{t^{\mu}},X_L=\frac{n_L}{t^{\mu}};\tau_M \right)
\nonumber 
\label{premeasureL}
\end{eqnarray}
with the scaling function
\begin{eqnarray}
D_{M,L_1}(X,X_L;\tau_M) =  \theta(t<\tau_M) \theta(0<X<X_L) 
 \frac{\mu}{\tau_M} \left(\frac{t}{\tau_M} \right)^{\mu}
e^{- X_L  }  
\label{measureL} 
\end{eqnarray}

In particular, the distribution of the trapping time $\tau_M$
is obtained, as it should, by simply normalizing the original distribution $q(\tau)$
on the interval $[t,+\infty[$
\begin{eqnarray}
q_t(\tau_M)= \int dX \int dX_L D_{M,L_1}(X,X_L;\tau_M) 
= \theta(\tau_M-t) \frac{\mu t^{\mu} }{\tau_M^{1+\mu}}
\label{tauM}  
\end{eqnarray}
The probability $\pi_{L_1} =\left( 1-e^{- \frac{t}{\tau_M} } \right)$
to have already escaped from the main trap at time $t$
and to be thus already in the trap $L_1$ reads
after averaging over the disorder
\begin{eqnarray}
\overline{ \pi_{L_1} }&&  
  = \int d\tau q_t(\tau) \left( 1-e^{- \frac{t}{\tau} } \right) 
 = \mu \int_0^1 dv v^{\mu-1} (1-e^{-v}) \\
&& =   \mu \int_0^1 \frac{dv}{v}  (1-e^{-v})
+  \mu^2 \int_0^1 \frac{dv}{v} \ln v (1-e^{-v}) +O(\mu^3)
\label{pil1}
\end{eqnarray}
so it is of order $\mu$.


At this level of approximation, the diffusion front
for a given sample is made out of two delta distributions
\begin{eqnarray}
P^{(0)+(1)}_{ML_1}(n)=  e^{- \frac{t}{\tau_M} } \delta_{n,n_M}
+ \left( 1-e^{- \frac{t}{\tau_M} } \right) \delta_{n,n_{L_1}}
\label{frontL1}
\end{eqnarray}

\subsection{ Statistical properties of the trap $S_1$}

\label{secs1}

The trap $S_1$ has been defined as the biggest trap before the main trap
$M$.
The joint distribution of the position $n$ of the main trap,
the position $n_S$ and the trapping time $\tau_S$ of the trap $S_1$ read
\begin{eqnarray}
{\cal D}_{S_1,M}(n_S,n;\tau_S) && = \theta(t>\tau_S) \left[ 1- \int_{\tau_s}^{+\infty} d\tau q(\tau) \right]^{n-1} q(\tau_s) \int_t^{+\infty} d\tau q(\tau) \\
&& \opsimeq_{t \to \infty} 
 \opsimeq_{t \to \infty} \frac{1}{t^{2 \mu} }
D_{S_1,M} \left(X_S=\frac{n_S}{t^{\mu}},X=\frac{n}{t^{\mu}};\tau_S \right)  
\label{premeasureS} 
\end{eqnarray}
where the scaling function reads
\begin{eqnarray}
D_{S_1,M}(X_S,X;\tau_S) \simeq  \theta(t>\tau_S>1) \theta(X>X_S> 0)  \frac{\mu}{\tau_S} \left(\frac{t}{\tau_S} \right)^{\mu}
e^{- X \left(\frac{t}{\tau_S} \right)^{\mu} }  
\label{measureS} 
\end{eqnarray}
We note that here, there are correlations between the trapping time 
and the positions, contrary to the decoupled measure (\ref{measureL})
concerning the trap $L_1$.
The joint distribution of the positions alone reads
\begin{eqnarray}
{\cal S}(X_s,X)  = \int d\tau_S D_{S_1,M}(X_S,X;\tau_S)
= \frac{ \theta(X>X_S> 0) }{X}  e^{-X}   
\end{eqnarray}
i.e. $X$ is distributed with $P^{(0)}=e^{-X}$ (\ref{gM}),
and $X_S$ is uniformly distributed on the interval $[0,X]$.
The distribution of the trapping time $\tau_s $ alone reads
\begin{eqnarray}
r(\tau_S)= \int_0^{+\infty} dX_S \int_{X_S}^{+\infty} dX D_{S_1,M}(X_S,X;\tau_S) =  \theta(t>\tau_S)   
\mu \frac{\tau_S^{\mu-1}}{t^{\mu}}  
\label{lawtauS}   
\end{eqnarray}
The probability $\pi_{S_1}=e^{-t/\tau_S}$ to be still in the trap $S_1$
at time $t$ reads
after averaging over the disorder 
\begin{eqnarray}
\overline{\pi_{S_1} } &&  
=\int d\tau_S r(\tau_S) e^{-\frac{t}{\tau_S} }
= \mu \int_1^{+\infty} \frac{dv}{v^{1+\mu} } e^{-v} \\
&& =  \mu \int_1^{+\infty} \frac{dv}{v } e^{-v}
- \mu^2 \int_1^{+\infty} \frac{dv}{v } e^{-v}+O(\mu^3)
\end{eqnarray}
so it is of order $\mu$.

At this level of approximation, the probability distribution reads
\begin{eqnarray}
P^{(0)+(1)}_{S_1M}(n)= e^{- \frac{t}{\tau_S} } \delta_{n,n_{S_1}}
+ \left( 1-e^{- \frac{t}{\tau_S} } \right) \delta_{n,n_M}
\label{frontS1}
\end{eqnarray}

We now use the statistical properties of the traps $L_1$ and $S_1$
to compute various observables at order $\mu$.

\subsection{ Diffusion front at order $\mu$ }

The correction due to the trap $L_1$
to the diffusion front in a given sample (\ref{frontL1})
with respect
to one delta function at the zero-th order (\ref{frontM}) reads
\begin{eqnarray}
P^{(1)}_{ML_1}(n) \equiv P^{(0)+(1)}_{ML_1}(n)-P^{(0)}(n)=   \left( 1-e^{- \frac{t}{\tau_M} } \right) (\delta_{n,n_{L_1}}-\delta_{n,n_M})
\label{correcfrontL1}
\end{eqnarray}

The average over the samples, that is
over the positions $(n_M,n_L)$ 
and over the trapping time $\tau$ 
with the measure (\ref{measureL}) yields 
the correction to the scaling function (\ref{defgfront}) 
\begin{eqnarray}
 g^{(1)}_{ML_1}(Y) &&  = \int d\tau_M \int dX \int dX_L
D_{M,L_1}(X,X_L;\tau_M) 
 \left( 1-e^{- \frac{t}{\tau_M} } \right) (\delta(Y-X_L)-\delta(Y-X))  \\
&& = e^{-X} (X-1) 
 \mu \int_0^1 dv v^{\mu-1}  (1-e^{-v}) 
 \\
&& = e^{-X} (X-1) \left[
 \mu \int_0^1 \frac{dv}{v}  (1-e^{-v})
+  \mu^2 \int_0^1 \frac{dv}{v} \ln v (1-e^{-v}) +O(\mu^3) \right]
\label{correcgL1} 
\end{eqnarray}

Similarly, the correction due to the trap $S_1$
to the diffusion front in a given sample (\ref{frontS1}) 
with respect
to one delta function at the zero-th order (\ref{frontM}) reads
\begin{eqnarray}
P^{(1)}_{S_1M}(n) \equiv P^{(0)+(1)}_{S_1M}(n)-P^{(0)}(n)= e^{- \frac{t}{\tau_S} } ( \delta_{n,n_{S_1}} 
- \delta_{n,n_M} ) 
\label{correcfrontS1}
\end{eqnarray}

After averaging over the samples
with the measure (\ref{measureS}), we obtain
the correction to the scaling function (\ref{defgfront}) 
\begin{eqnarray}
g^{(1)}_{S_1M}(Y) && 
= \int d\tau_S \int dX \int dX_S D_{S_1,M}(X_S,X;\tau_S)
 e^{-\frac{t}{\tau_S} }
 [ \delta(Y-X_S) -\delta( Y-X) ]   \\
 && =   \mu   \int_1^{+\infty} \frac{dv}{v} e^{-v}(1-Y  v^{\mu})  
e^{- Y v^{\mu} } \\  
&& =  \mu e^{-X} (1-X)   \int_1^{+\infty} \frac{dv}{v} e^{-v}
+ \mu^2 e^{-X} (X^2-2 X)  \int_1^{+\infty} \frac{dv}{v} \ln v e^{-v} +O(\mu^3)
\label{correcgS1} 
\end{eqnarray}

Adding these two contributions to the zeroth order front (\ref{gM}),
we finally get 
\begin{eqnarray}
g^{(0)+(1)}_{total} \equiv g^{(0)}(X)+g^{(1)}_{ML_1}(X) + g^{(1)}_{S_1M}(X)
  =  e^{-X}+ e^{-X} (X-1) \mu \gamma_E
 +O(\mu^2)
\end{eqnarray}
which coincides
with the expansion at order $\mu$ of the exact Levy front
(\ref{fmu}).

\subsection{ Thermal width at order $\mu$ }

For a given sample, the contribution of the trap $L_1$
to the thermal width reads (\ref{frontL1})
\begin{eqnarray}
\left[<\Delta n^2(t)> \right]^{(1)}_{ML_1} &&  = \overline{ <n^2> - <n>^2 }
=  e^{-  \frac{t}{\tau_M}} (1-e^{-  \frac{t}{\tau_M} } )
(n_L-n_M)^2 
 \end{eqnarray}
Averaging over the disorder, that is over the positions 
and over the trapping time $\tau$ 
with the measure (\ref{measureL}) yields :
\begin{eqnarray}
\left[ \Delta(\mu) \right]^{(1)}_{ML_1}
= \left[ \frac{ \overline{ <\Delta n^2(t)> } } {t^{2 \mu} }  
\right]^{(1)}_{ML_1}
&& = 2  \mu \int_0^1 dv v^{\mu-1}
 e^{-  v }   (1-e^{-  v } ) \\
&& = 2   \mu \int_0^1 \frac{dv}{v}
 e^{-  v }   (1-e^{-  v } )
+ 2   \mu^2 \int_0^1 \frac{dv}{v} \ln v
 e^{-  v }   (1-e^{-  v } )
+O(\mu^3) 
\label{thermalL1}
 \end{eqnarray}

Similarly, the contribution of the trap $S_1$ (\ref{frontS1})
averaged over the samples with the measure (\ref{measureS})
reads
\begin{eqnarray}
\left[ \Delta(\mu) \right]^{(1)}_{S_1M}
=\left[ \frac{ \overline{ <\Delta n^2(t)> } } {t^{2 \mu} }  
\right]^{(1)}_{S_1M}  && = \overline{ e^{-  \frac{t}{\tau_S}} (1-e^{-  \frac{t}{\tau_s} } )
\frac{ (n_L-n_s)^2  }{t^{2 \mu}} }  = 2  \mu
\int_1^{+\infty} dv  v^{-1- 3 \mu} e^{-v} (1-e^{-v}) \\
&& =  2   \mu
\int_1^{+\infty} \frac{dv}{v} e^{-v} (1-e^{-v})
- 6   \mu^2
\int_1^{+\infty} \frac{dv}{v} \ln v e^{-v} (1-e^{-v}) +O(\mu^3) 
\label{thermalS1}
\end{eqnarray}

Adding the two contributions finally yields
\begin{eqnarray}
\left[ \Delta(\mu) \right]^{(1)}_{total}
 =    2 \mu \int_0^{+\infty} \frac{dv}{v}
 e^{-  v }   (1-e^{-  v } ) 
=   \mu (2 \ln 2)   
\end{eqnarray}
in agreement with the exact result (\ref{widthexact}) of \cite{aslangul}.

\subsection{ Localization parameters at order $\mu$ }

For a given sample, the contribution of the trap $L_1$ (\ref{frontL1})
to the localization parameter $Y_k$ representing
the probability to find $k$ independent particles in the same trap
at time $t$ reads
\begin{eqnarray}
[Y_k]^{(0)+(1)}_{ML_1}(t) && = \left( e^{-  \frac{t}{\tau_M}} \right)^k
+  \left( 1-e^{-  \frac{t}{\tau_M} } \right)^k 
\end{eqnarray}
So after averaging over the samples,
that is over the trapping time $\tau$ (\ref{tauM}),
the correction to the zero-th order (\ref{ykm}) due to the trap $L_1$ reads
\begin{eqnarray}
\overline{[Y_k]^{(1)}_{ML_1}(t)  } && \equiv
\overline{[Y_k]^{(0)+(1)}_{ML_1}(t) -Y_k^{(0)}(t) }
 = \mu \int_0^1 dv v^{\mu-1} 
\left[ e^{- k v }   + (1-e^{-  v } )^k -1 \right]
\label{correcYkL1}
\end{eqnarray}

Similarly, the correction to the zero-th order (\ref{ykm}) due to the trap $S_1$ reads after averaging over the samples, that is over the trapping time
$\tau_S$ (\ref{lawtauS}) 
\begin{eqnarray}
\overline{[Y_k]^{(1)}_{S_1M}(t)  } \equiv
\overline{[Y_k]^{(0)+(1)}_{S_1M}(t) -Y_k^{(0)}(t) }
 = \mu \int_1^{+\infty} dv v^{-\mu-1}  
[ e^{- k v  }
+  (1-e^{-  v } )^k -1 ]
\label{correcYkS1}
\end{eqnarray}

Adding these two contributions, we finally get at first order in $\mu$
\begin{eqnarray}
\overline{Y_k^{(0)+(1)}} \equiv \overline{Y_k^{(0)}} + 
\overline{[Y_k]^{(1)}_{ML_1}(t) }
+\overline{[Y_k]^{(1)}_{S_1M}(t) } = 1+  \mu \int_0^{+\infty} \frac{dv}{v}  
[ e^{- k v  }
+  (1-e^{-  v } )^k -1 ] +O(\mu^2)
\end{eqnarray}
For the special case $k=2$, the result 
\begin{eqnarray}
 \overline{Y_2^{(0)+(1)}} = 1- (2 \ln 2) \mu  +O(\mu^2)
\label{specialy2}
\end{eqnarray}
is again in agreement with the expansion  (\ref{y2exact})
of the exact result \cite{comptejpb}.
The other first values of $k$ yield
\begin{eqnarray}
 \overline{Y_3^{(0)+(1)}}  && = 1- (3 \ln 2) \mu  +O(\mu^2)  \\
 \overline{Y_4^{(0)+(1)}}  && = 1- (2 \ln \frac{32}{9}) \mu  +O(\mu^2) \end{eqnarray}

\subsection{ Correlation function of two particles 
at order $\mu$ }

We now consider the correlation function of two particles (\ref{correform}).
For a given sample, the contribution of the trap $L_1$ (\ref{frontL1})
reads
\begin{eqnarray}
 C^{(0)+(1)}_{ML_1}(l,t) && \equiv \sum_{n=0}^{+\infty} \sum_{m=0}^{+\infty}
P^{(0)+(1)}_{ML_1}(n) P^{(0)+(1)}_{ML_1}(m) \delta_{l,\vert n-m \vert} \\
&& = [ \left( e^{-  \frac{t}{\tau_M}} \right)^2
+  \left( 1-e^{-  \frac{t}{\tau_M} } \right)^2 ] \delta_{l,0}
+ 2  e^{-  \frac{t}{\tau_M}} \left( 1-e^{-  \frac{t}{\tau_M} } \right)
\delta_{l,(n_{L_1}-n_M)} 
\end{eqnarray}
After averaging over the disorder with the measure (\ref{measureL}), the correction 
with respect to the zero-th order of the correlation function (\ref{correm})
reads
\begin{eqnarray}
\overline{  C^{(1)}_{ML_1}(l,t)  }
\equiv \overline{  C^{(0)+(1)}_{ML_1}(l,t) - C^{(0)}(l,t) }
\opsimeq \overline{ [Y_2]^{(1)}_{ML_1} } \delta_{l,0} 
+ 2 \mu \frac{1}{t^{\mu}} e^{- \frac{l}{t^{\mu}} }
 \int_0^1 dv v^{\mu-1}
 e^{-  v }   (1-e^{-  v } )  \end{eqnarray}
It presents the form (\ref{correform}) : the weight of the $\delta$ part has been obtained in (\ref{correcYkL1}) and the scaling function reads
\begin{eqnarray}
\left[ {\cal C}_{\mu}(\lambda) \right]_{ML_1} && =
e^{- \lambda } \  2 \mu \int_0^1 dv v^{\mu-1}
 e^{-  v }   (1-e^{-  v } )   \\
&&= e^{- \lambda } 
\left[ 2 \mu \int_0^1 \frac{dv}{v} 
 e^{-  v }   (1-e^{-  v } )
+ 2 \mu^2 \int_0^1 \frac{dv}{v}  \ln v
 e^{-  v }   (1-e^{-  v } ) +O(\mu^3) \right]  
\label{correML1}
\end{eqnarray}

Similarly, the contribution of the trap $S_1$leads
after averaging over the samples with the measure (\ref{measureS})
\begin{eqnarray}
\overline{  C^{(1)}_{S_1M}(l,t)  }
\equiv\overline{  C^{(0)+(1)}_{S_1M} - C^{(0)}(l,t) }
&& = \overline{ [ \left( e^{-  \frac{t}{\tau_S}} \right)^2
+  \left( 1-e^{-  \frac{t}{\tau_S} } \right)^2 -1 ] \delta_{l,0}
+ 2  e^{-  \frac{t}{\tau_S}} \left( 1-e^{-  \frac{t}{\tau_S} } \right)
\delta_{l,(n_{M}-n_{S_1})}}   \nonumber
\\
&& \opsimeq \overline{ [Y_2]^{(1)}_{S_1M} } \delta_{l,0} 
+ 2 \mu \int_1^{+\infty} \frac{dv}{v} 
 e^{-  v }   (1-e^{-  v } ) 
 \frac{1}{t^{\mu}} e^{- \frac{l}{t^{\mu}} v^{\mu} }
\end{eqnarray}
It presents the form (\ref{correform}) : the weight of the $\delta$ part has been obtained in (\ref{correcYkS1}) and the scaling function reads
\begin{eqnarray}
\left[ {\cal C}_{\mu}(\lambda) \right]_{S_1M}=
&&  2 \mu \int_1^{+\infty} \frac{dv}{v} 
 e^{-  v }   (1-e^{-  v } ) 
  e^{- \lambda v^{\mu} }\\
= && 
e^{- \lambda   } \left[ 2 \mu \int_1^{+\infty} \frac{dv}{v} 
 e^{-  v }   (1-e^{-  v } ) 
- 2 \mu^2 \lambda \int_1^{+\infty} \frac{dv}{v} \ln v
 e^{-  v }   (1-e^{-  v } ) 
     \right]
\label{correS1M}
\end{eqnarray}

Adding these two contributions, we finally get 
at first order in $\mu$ the following correlation function
\begin{eqnarray}
\overline{C^{(0)+(1)}(l,t)} && =\overline{C^{(0)}(l,t)}
+\overline{ C^{(1)}_{ML_1} }
+ \overline{ C^{(1)}_{S_1M} } \nonumber \\
&& \opsimeq  \left[ 1 - (2 \ln 2) \mu  +O(\mu^2) \right] \delta_{l,0} 
 + \frac{1}{t^{\mu}} e^{- \frac{l}{t^{\mu}} }
\left[ (2 \ln 2) \mu   +O(\mu^2) \right]
\end{eqnarray}

\subsection{ Generating function of thermal cumulants  at first order in $\mu$ }

The correction to the generating function (\ref{gencum})
due to the trap $L_1$ (\ref{frontL1}) with respect to the zero-th order
(\ref{gen0}) reads with the measure (\ref{measureL}) 
\begin{eqnarray}
[ Z_{\mu}(s) ]^{(1)}_{ML_1} && \equiv 
[ Z_{\mu}(s) ]^{(0)+(1)}_{ML_1}- Z_{\mu}^{(0)}(s) 
 = \overline{ \ln \left[ e^{- \frac{t}{\tau_M}}
+(1-e^{-\frac{t}{\tau_M}}) e^{-s (X_L-X) } \right] } \nonumber \\
&& = \mu \int_0^1 dv v^{\mu-1} \int_0^{+\infty} dY
e^{-Y}  \ln \left[e^{- v}
+(1-e^{- v}) e^{-s Y } \right] 
\label{zmuML1}
\end{eqnarray}

Similarly, the correction due to the trap $S_1$ reads
with the measure (\ref{measureS})
\begin{eqnarray}
[ Z_{\mu}(s) ]^{(1)}_{S_1M} && \equiv 
[ Z_{\mu}(s) ]^{(0)+(1)}_{S_1M}- Z_{\mu}^{(0)}(s) 
 = \overline{ \ln \left[e^{- \frac{t}{\tau_s}} e^{ s (X-X_S)
+(1-e^{-\frac{t}{\tau_s}})  }  \right] } \nonumber   \\
&& = \mu \int_1^{+\infty} \frac{dv}{v} \int_0^{+\infty} dY
e^{-Y v^{\mu} }  \ln \left[e^{- v} e^{sY}
+(1-e^{- v}) \right]
\label{zmuS1M}
\end{eqnarray}

The total correction at order $\mu$ thus reads
\begin{eqnarray}
[ Z_{\mu}(s) ]^{(1)}_{total} && =[ Z_{\mu}(s) ]^{(1)}_{ML_1}+[ Z_{\mu}(s) ]^{(1)}_{S_1M}  \nonumber \\
&& = \mu \int_0^{+\infty} dY
e^{-Y} \left[ \int_0^1 \frac{dv}{v}    \ln \left[e^{- v}
+(1-e^{- v}) e^{-s Y } \right]
+ \int_1^{+\infty} \frac{dv}{v}   \ln \left[e^{- v} e^{sY}
+(1-e^{- v}) \right] \right]
\end{eqnarray}
We may now perform a series expansion in $s$
and evaluate the integrals to obtain
the generating function of all thermal cumulants
at first order in $\mu$
\begin{eqnarray}
[ Z_{\mu}(s) ]^{(1)}_{total} && = -s \gamma_E+s^2 \ln 2 - s^3 
(2 \ln 3 -3 \ln 2) +  s^4(19 \ln 2 - 12 \ln 3) +O(s^5)
\end{eqnarray}
leading to the results (\ref{cumultrap}).

\section{ Directed trap model at order $\mu^2 $ }

\label{trapmu2}

\subsection{Dispersion of the thermal packet at order $\mu^2$ }

To compute observables at order $\mu^2$, we now have to consider
the possible dispersions of the thermal packet over
three traps. 
Denoting by $\tau_1$ and $\tau_2$ the first two trapping times,
the occupation probabilities of the three ordered sites are given by
\begin{eqnarray}
p_1(t;\tau_1) && = e^{- \frac{t}{\tau_1}} \nonumber \\
p_2(t;\tau_1,\tau_2) && = \frac{\tau_2}{\tau_2-\tau_1} 
\left( e^{-\frac{t}{\tau_2}}-e^{-\frac{t}{\tau_1}}\right) \nonumber \\
p_3(t;\tau_1,\tau_2) && = 1- \frac{\tau_2 e^{-\frac{t}{\tau_2}} - \tau_1 e^{-\frac{t}{\tau_1}}}{\tau_2-\tau_1} 
\label{occ3}
\end{eqnarray}
In the following, we will also use the notations $v=\frac{t}{\tau_1}$,
$w=\frac{t}{\tau_2}$ 
\begin{eqnarray}
p_1(v) && = e^{- v } \nonumber \\
p_2(v,w) && = \frac{v}{v-w} ( e^{-w}-e^{-v}) \nonumber \\
p_3(v,w) && = 1- p_1(v) -p_2(v,w) 
\label{occvw}
\end{eqnarray}
To simplify computations later, it will be convenient to use 
in intermediate calculations the two following
obvious properties : 
the occupation probability of the third site is 
a symmetric function of $(v,w)$
\begin{eqnarray}
p_3(v,w)  = p_3(w,v) 
\label{sym3}
\end{eqnarray}
and the three occupation probabilities satisfy the normalization
\begin{eqnarray}
p_1(v) +
p_2(v,w) +
p_3(v,w) =1 
\label{norma3}
\end{eqnarray}

The enumeration of the various possibilities for the three traps
is as follows (see Figure \ref{trap}).

\subsubsection{ Configurations $(M,L_1,L_2)$}

The tree traps are the main trap M, the next renormalized trap $L_1$
introduced in (\ref{secl1}) and the second next renormalized trap 
that we call $L_2$.
The joint distribution of the rescaled positions and trapping times read
\begin{eqnarray}
T_{M,L_1,L_2} (X,X_{ML_1},X_{L_2} ;\tau_M,\tau_{L_1} )
 = && \theta(t< \tau_M) \theta(t<\tau_{L_1} )\theta(0 \leq X \leq X_1\leq X_2 )  \nonumber \\
&&  \frac{\mu}{\tau_M} \left(\frac{t}{\tau_M} \right)^{\mu}
\frac{\mu}{\tau_{L_1}} \left(\frac{t}{\tau_{L_1}} \right)^{\mu}
e^{-X_2}
\label{tML1L2}
\end{eqnarray}
At this level of approximation, the diffusion front is made out of
three delta pics
\begin{eqnarray}
P^{(0)+(1)+(2)}_{ML_1L_2} ( n )
= p_1(t;\tau_M) \delta_{m,n_M} + p_2(t;\tau_M,\tau_{L_1}) \delta_{m,n_{L_1}} + p_3(t;\tau_M,\tau_{L_1}) \delta_{m,n_{L_2}}
\label{frontML1L2}
\end{eqnarray}
where the weights of the three traps
are given by (\ref{occ3}).

\subsubsection{ Configurations $(M,I_2,L_1)$}

The tree traps are the main trap M, the next renormalized trap $L_1$
introduced in (\ref{secl1}) and in between 
the intermediate trap 
that we call $I_2$, defined as the biggest trap
in the decimated region between M and $L_1$.
The joint distribution of the rescaled positions and trapping times read
\begin{eqnarray}
 T_{M,I_2,L_1}(X,X_I,X_L;\tau_M,\tau_I) 
&&=  \theta(\tau_M>t>\tau_I>1) \theta(X_L>X_I>X> 0) 
\nonumber  \\
&& \frac{\mu}{\tau_M} \left(\frac{t}{\tau_M} \right)^{\mu}
 \frac{\mu}{\tau_I} \left(\frac{t}{\tau_I} \right)^{\mu}
e^{-X} e^{- (X_L-X) \left(\frac{t}{\tau_I} \right)^{\mu} }  
\label{tMI1L1}
\end{eqnarray}
The corresponding diffusion front reads
\begin{eqnarray}
P^{(0)+(1)+(2)}_{MI_2L_1} ( n )
= p_1(t;\tau_M) \delta_{m,n_M} + p_2(t;\tau_M,\tau_I) \delta_{m,n_{I_2}} + p_3(t;\tau_M,\tau_I) \delta_{m,n_{L_1}}
\label{frontMI1L1}
\end{eqnarray}
where the weights 
are given by (\ref{occ3}).

\subsubsection{   Configurations $(S_1,S_2,M)$ and $(S_2',S_1,M)$}

The tree traps are the main trap M, the trap $S_1$ defined as before
as the biggest trap before $M$, and the second biggest trap before $M$,
that we call $S_2$ if its position is between $S_1$ and $M$,
and that we call $S_2'$ if its position is between $0$ and $S_1$.

For the configurations $(S_1,S_2,M)$, the joint distribution 
of the rescaled positions and trapping times is given by
\begin{eqnarray}
 T_{S_1,S_2,M}(X_1,X_2,X;\tau_{S_1},\tau_{S_2}) 
&&=  \theta(\tau_{S_2}<\tau_{S_1}<t)  \theta(0<X_1<X_2<X)  
\nonumber \\
&& \frac{\mu}{\tau_{S_1}} \left(\frac{t}{\tau_{S_1}} \right)^{\mu}
 \frac{\mu}{\tau_{S_2}} \left(\frac{t}{\tau_{S_2}} \right)^{\mu}
 e^{- X \left(\frac{t}{\tau_{S_2}} \right)^{\mu} } 
\label{tS1S2M}
\end{eqnarray}
and the corresponding diffusion front reads
\begin{eqnarray}
P^{(0)+(1)+(2)}_{S_1S_2M} ( n )
= p_1(t;\tau_{S_1}) \delta_{m,n_{S_1}} + p_2(t;\tau_{S_1},\tau_{S_2}) \delta_{m,n_{S_2}} + p_3(t;\tau_{S_1},\tau_{S_2}) \delta_{m,n_M}
\label{frontS1S2M}
\end{eqnarray}
where the weights 
are given by (\ref{occ3}).

For the configurations $(S_2',S_1,M)$
the joint distribution 
of the rescaled positions and trapping times read
\begin{eqnarray}
 T_{S_2',S_1,M}(X_2,X_1,X;\tau_{S_2},\tau_{S_1},) 
&& =  \theta(\tau_{S_2}<\tau_{S_1}<t)  \theta(0<X_2<X_1<X) 
\nonumber  \\
&& \frac{\mu}{\tau_{S_1}} \left(\frac{t}{\tau_{S_1}} \right)^{\mu}
 \frac{\mu}{\tau_{S_2}} \left(\frac{t}{\tau_{S_2}} \right)^{\mu}
 e^{- X \left(\frac{t}{\tau_{S_2}} \right)^{\mu} } 
\label{tS2pS1M}
\end{eqnarray}
and the corresponding diffusion front reads
\begin{eqnarray}
P^{(0)+(1)+(2)}_{S_2'S_1M} ( n )
= p_1(t;\tau_{S_2}) \delta_{m,n_{S_2'}} + p_2(t;\tau_{S_2},\tau_{S_1}) \delta_{m,n_{S_1}} + p_3(t;\tau_{S_2},\tau_{S_1}) \delta_{m,n_M}
\label{frontS2S1M}
\end{eqnarray}
where the weights 
are given by (\ref{occ3}).

\subsubsection{   Configurations $(S_1,M,L_1)$}

The three traps are the trap $S_1$introduced in (\ref{secs1}),
the main trap M and the next renormalized trap $L_1$
introduced in (\ref{secl1}).
The joint distribution of the rescaled positions and trapping times 
is given by
\begin{eqnarray}
 T_{S_1,M,L_1}(X_s,X,X_L;\tau_S,\tau_M) 
&& =  \theta(\tau_M>t>\tau_S>1) \theta(X_L>X>X_S> 0) 
\nonumber \\
&&  \frac{\mu}{\tau_S} \left(\frac{t}{\tau_S} \right)^{\mu}
 \frac{\mu}{\tau_M} \left(\frac{t}{\tau_M} \right)^{\mu}
 e^{- X \left(\frac{t}{\tau_S} \right)^{\mu} }  e^{-(X_L-X)} 
\label{tS1ML1}
\end{eqnarray}
The corresponding diffusion front reads
\begin{eqnarray}
P^{(0)+(1)+(2)}_{S_1ML_1} ( n )
= p_1(t;\tau_{S_1}) \delta_{m,n_{S_1}} + p_2(t;\tau_{S_1},\tau_M) \delta_{m,n_M} + p_3(t;\tau_{S_1},\tau_M) \delta_{m,n_{L_1}}
\label{frontS1ML1}
\end{eqnarray}
where the weights 
are given by (\ref{occ3}).

We now use the statistical properties of these three-traps configurations
to compute observables at order $\mu^2$.

\subsection { Diffusion front at order $\mu^2$ }

\subsubsection{ Contributions at order $\mu^2$ of the two-traps configurations }

We have already studied the contributions of two traps configurations
when studying the order $\mu$.
The contribution of order $\mu^2$ of the configurations $ML_1$ (\ref{correcgL1})
reads
\begin{eqnarray}
 g^{(2)}_{ML_1}(X)  && =  \mu^2 e^{-X} (X-1) 
 \int_0^{1} \frac{dv}{v } \ln v (1-e^{-v})
\label{frontML1mu2}
\end{eqnarray}
Similarly, the contribution of order $\mu^2$ of the configurations $S_1M$ (\ref{correcgS1})
reads
\begin{eqnarray}
g^{(2)}_{S_1M}(X)  && =
 \mu^2 e^{-X}  (X^2 - 2 X) 
 \int_1^{+\infty} \frac{dv}{v} \ln v e^{-v} 
\label{frontS1Mmu2}
\end{eqnarray}

\subsubsection{ Contributions 
at order $\mu^2$ of the three-traps configurations }

The specific contribution at order $\mu^2$ of the three-traps configurations
of type $ML_1L_2$ can be obtained by subtracting from (\ref{frontML1L2})
the two-traps configurations $ML_1$ (\ref{frontL1})
\begin{eqnarray}
 P^{(2)}_{ML_1L_2} ( n ) 
&& \equiv  P^{(0)+(1)+(2)}_{ML_1L_2} ( n )
- P^{(0)+(1)+(2)}_{ML_1} ( n )  =  p_3(t;\tau_M,\tau_{L_1}) 
\left(\delta_{m,n_{L_2}} - \delta_{m,n_{L_1}} \right) 
\end{eqnarray}
The average over the samples with the measure (\ref{tML1L2}) yields
the correction of the scaling function (\ref{defgfront})
\begin{eqnarray}
 g^{(2)}_{ML1L2} (Y) &&
  = \int dX \int dX_1 \int dX_2 \int \tau_M \int \tau_{L_1}
\ \ T_{M,L_1,L_2} (X,X_{ML_1},X_{L_2} ;\tau_M,\tau_{L_1} ) \\
&& p_3(t;\tau_M,\tau_{L_1}) 
\left(\delta(Y-X_2) - \delta(Y-X_1) \right) \\
&& = e^{- Y} \left[  \frac{Y^2}{2}-Y  \right] \mu^2 \int_0^1 \frac{dv}{v} \int_0^1 \frac{dw}{w} p_3(v,w)+O(\mu^3) 
 \end{eqnarray}

Similarly, the specific contribution at order $\mu^2$ of the three-traps configurations
of type $MI_2L_1$ can be obtained by subtracting from (\ref{frontMI1L1})
the two-traps configurations $ML_1$ (\ref{frontL1})
and this yields after averaging over the samples
 with the measure (\ref{tMI1L1})
\begin{eqnarray}
\overline{ P^{(2)}_{MI_2L_1} ( n ) }
&& \equiv \overline{ P^{(0)+(1)+(2)}_{MI_2L_1} ( n )
- P^{(0)+(1)+(2)}_{ML_1} ( n ) } =
\overline{ p_2(t;\tau_M,\tau_I)  
\left(\delta_{m,n_{I_2}} - \delta_{m,n_{L_1}} \right) } 
\end{eqnarray}
The correction to the scaling function (\ref{defgfront}) thus reads
\begin{eqnarray}
 g^{(2)}_{MI_2L_1} (Y)  =  e^{- Y} \left[ Y- \frac{Y^2}{2}  \right]
\mu^2 \int_0^1 \frac{dv}{v} \int_1^{+\infty}  \frac{dw}{w}
 p_2(v,w) +O(\mu^3)
  \end{eqnarray}

For the three-traps configurations
of type $S_1S_2M$, we have to subtract from (\ref{frontS1S2M})
the two-traps configurations $S_1M$ (\ref{frontS1})
and to average over the samples
 with the measure (\ref{tS1S2M})
\begin{eqnarray}
 \overline{ P^{(2)}_{S_1S_2M} ( n ) }
&& \equiv \overline{ P^{(0)+(1)+(2)}_{S_1S_2M} ( n )
- P^{(0)+(1)+(2)}_{S_1M} ( n ) } =
\overline{  p_2(t;\tau_{S_1},\tau_{S_2}) 
(\delta_{m,n_{S_2}} - \delta_{m,n_M}) } 
\end{eqnarray}
The correction to the scaling function (\ref{defgfront}) thus reads
\begin{eqnarray}
 g^{(2)}_{S_1S_2M} (Y)  =  e^{- Y} \left[ Y-\frac{Y^2}{2}  \right]
\mu^2 \int_1^{+\infty} \frac{dv}{v} \int_v^{+\infty}  \frac{dw}{w}
 p_2(v,w) +O(\mu^3)
  \end{eqnarray}

For the three-traps configurations
of type $S_2'S_1M$, we have to subtract from (\ref{frontS2S1M})
the two-traps configurations $S_1M$ (\ref{frontS1})
and to average over the samples
 with the measure (\ref{tS2pS1M})
\begin{eqnarray}
&& \overline{ P^{(2)}_{S_2'S_1M} ( n ) }
 \equiv \overline{ P^{(0)+(1)+(2)}_{S_2'S_1M} ( n )
- P^{(0)+(1)+(2)}_{S_1M} ( n ) } = \\
&& \overline{ p_1(t;\tau_{S_2}) \delta_{m,n_{S_2'}} 
+ (p_2(t;\tau_{S_2},\tau_{S_1}) - p_2(t;0,\tau_{S_1}))
\delta_{m,n_{S_1}} + (p_3(t;\tau_{S_2},\tau_{S_1})
- p_3(t;0,\tau_{S_1})) \delta_{m,n_M} } 
\end{eqnarray}
The correction to the scaling function (\ref{defgfront}) reads
\begin{eqnarray}
 g^{(2)}_{S_2'S_1M} (Y)
= e^{- Y} (1-Y) \mu^2 \int_1^{+\infty} \frac{dv}{v} \ln v e^{-v}   
+  e^{- Y} (Y-\frac{Y^2}{2}) \mu^2 \int_1^{+\infty} \frac{dv}{v} \int_1^{+\infty}  \frac{dw}{w} \theta(v<w) [  
 p_2(v,w)] +O(\mu^3)
  \end{eqnarray}

For the three-traps configurations
of type $S_1ML_1$, we have to subtract from (\ref{frontS1ML1})
the one trap configuration (\ref{frontM}),
and the corrections due to the two-traps configurations 
$ML_1$ (\ref{correcfrontL1})and $S_1M$ (\ref{correcfrontS1}),
and to average over the samples
 with the measure (\ref{tS1ML1})
\begin{eqnarray}
\overline{ P^{(2)}_{S_1ML_1} ( n ) }
&& \equiv \overline{ P^{(0)+(1)+(2)}_{S_1ML_1} ( n )
- P^{(0)}_{M} ( n ) - P^{(1)+(2)}_{S_1M} ( n )
- P^{(1)+(2)}_{ML_1} ( n )} \nonumber  \\
&& = \overline{ [p_3(t;\tau_{S_1},\tau_M)- p_3(t;0,\tau_M)]
( \delta_{m,n_{L_1}} - \delta_{m,n_M} ) } 
\end{eqnarray}
with the scaling function
\begin{eqnarray}
 g^{(2)}_{S1ML1} (Y)  =  e^{- Y} \left[ Y-\frac{Y^2}{2}  \right]
\mu^2 \int_1^{+\infty} \frac{dw}{w} \int_0^{1} \frac{dv}{v} 
p_2(v,w)   
+O(\mu^3)
 \end{eqnarray}

The sum of all contributions of order $\mu^2$ finally reads
\begin{eqnarray}
g^{(2)}_{ \rm{total} } (Y) 
&& \equiv g^{(2)}_{MI_2L_1} (Y) + g^{(2)}_{S_1ML_1} (Y)
+ g^{(2)}_{S_1S_2M} (Y) + g^{(2)}_{S_2'S_1M} (Y)+ g^{(2)}_{S_1M}(Y)
+ g^{(2)}_{ML1L2} (Y)  +  g^{(2)}_{ML_1}(Y) \\
&& =
 e^{- Y} \left[ 2 Y- Y^2  \right]
\mu^2  \int_0^{+\infty} \frac{dv}{v} \int_v^{+\infty} \frac{dw}{w}
p_2(v,w) \nonumber \\ && + \mu^2 e^{-Y}  (Y^2- 3Y+1) 
\left[ \int_1^{+\infty} \frac{dv}{v} \ln v e^{-v}
 - \int_0^1 \frac{dv}{v} \ln v ( 1- e^{-v}) \right] +O(\mu^3)
\end{eqnarray}

The double integral may be computed as follows
\begin{eqnarray}
 \int_0^{+\infty} \frac{dv}{v} \int_v^{+\infty} \frac{dw}{w}
p_2(v,w) =
    \int_0^{+\infty} \frac{dw}{w} \int_0^{1}  dz
\frac{1}{z-1} (e^{-wz}-e^{-w}) 
=     \int_0^{1}  dz
\frac{1}{1-z} (\ln z) = - \frac{\pi^2}{6} 
\label{double}
\end{eqnarray}
and we obtain the final result

\begin{eqnarray}
g^{(2)}_{ \rm{total} } (Y) 
 = \mu^2 e^{- Y} \left[ \left( \frac{\gamma_E^2}{2} + \frac{\pi^2}{12} \right)
 + Y \left( -3 \frac{\gamma_E^2}{2} + \frac{\pi^2}{12} \right) 
 + Y^2 \left( \frac{\gamma_E^2}{2} - \frac{\pi^2}{12} \right) \right] 
\end{eqnarray}
which coincides with the expansion (\ref{fmu}) of the exact diffusion
front described in Appendix (\ref{levy}).

\subsection { Thermal width at order $\mu^2$ }

\subsubsection{ Contributions at order $\mu^2$ of the two-traps configurations }

We have already studied the contributions of two traps configurations
when studying the order $\mu$.
The contribution of order $\mu^2$ of the configurations $ML_1$ 
reads (\ref{thermalL1})
\begin{eqnarray}
\left[ \Delta(\mu) \right]^{(2)}_{ML_1}
= 2   \mu^2 \int_0^{1} \frac{dv}{v} \ln v
 e^{-  v }   (1-e^{-  v } )
\label{thermalML1mu2}
 \end{eqnarray}
whereas the contribution of the configurations $S_1M$
reads (\ref{thermalS1})
\begin{eqnarray}
\left[ \Delta(\mu) \right]^{(2)}_{S_1M} 
= - 6   \mu^2
\int_1^{+\infty} \frac{dv}{v} \ln v e^{-v} (1-e^{-v})
\label{thermalS1Mmu2}
\end{eqnarray}

\subsubsection{ Contributions at order $\mu^2$ of
 the three-traps configurations }

For a given configuration of three traps situated at $(n_1,n_2,n_3)$
 with occupation probabilities $(p_1,p_2,p_3)$, the 
thermal width reads
\begin{eqnarray}
 <\Delta n^2(t)> = p_1 p_2 (n_2-n_1)^2 
+ p_1 p_3 (n_3-n_1)^2 + p_2 p_3 (n_3-n_2)^2
\end{eqnarray}

Following the procedure described above for the diffusion front,
we obtain the specific contributions at order $\mu^2$
of the various configurations as follows.


The configurations of type $ML_1L_2$  with the measure (\ref{tML1L2}) give
\begin{eqnarray}
&& \left[\Delta(\mu) \right]^{(2)}_{ML_1L_2}
 \equiv   \left[\Delta(\mu) \right]^{(1)+(2)}_{ML_1L_2}
- \left[\Delta(\mu)\right]^{(1)+(2)}_{ML_1}  
\\ &&  = 
\overline{   p_2(t;\tau_M,\tau_{L_1}) p_3(t;\tau_M,\tau_{L_1}) [  X_{L_2} - X_{L_1}  ]^2
+  p_1(t;\tau_M) p_3(t;\tau_M,\tau_{L_1})
 [ (X_{L_2}-X_M)^2- (X_{L_1}-X_M)^2  ] } 
\\   
&& = \mu^2 \int_0^1 \frac{dv}{v} \int_0^1  \frac{dw}{w}
\left[ 2  p_2(v,w) p_3(v,w) 
+  4 p_1(v) p_3(v,w) \right] +O(\mu^3)
 \end{eqnarray}



The configurations of type $MI_2L_1$ with the measure (\ref{tMI1L1}) give
\begin{eqnarray}
&& \left[\Delta(\mu) \right]^{(2)}_{MI_2L_1}
 \equiv  \left[\Delta(\mu) \right]^{(1)+(2)}_{MI_2L_1}
- \left[\Delta(\mu) \right]^{(1)+(2)}_{ML_1}  
\\ &&  = 
\overline{   p_2(t;\tau_M,\tau_{I_2}) p_3(t;\tau_M,\tau_{I_2})
 [  X_{L_1} - X_{I_2}  ]^2
+  p_1(t;\tau_M) p_2(t;\tau_M,\tau_{I_2}) 
[ (X_{I_2}-X_M)^2- (X_{L_1}-X_M)^2  ] } \\
&& = \mu^2 \int_0^1 \frac{dv}{v} \int_1^{+\infty}  \frac{dw}{w}
\left[ 2  p_2(v,w) p_3(v,w)
-4  p_1(v) p_2(v,w) \right]
 \end{eqnarray}


The configurations of type $S_1S_2M$ with the measure (\ref{tS1S2M}) give
\begin{eqnarray}
&&  \left[ \Delta(\mu) \right]^{(2)}_{S_1S_2M}
 \equiv  \left[\Delta(\mu) \right]^{(1)+(2)}_{S_1S_2M}
- \left[\Delta(\mu) \right]^{(1)+(2)}_{S_1M}  
\\ &&  = 
\overline{  p_2(t;\tau_{S_1},\tau_{S_2}) p_3(t;\tau_{S_1},\tau_{S_2})
 [  X_{M} - X_{S_2}  ]^2
+  p_1(t;\tau_{S_1}) p_2(t;\tau_{S_1},\tau_{S_2}) [ (X_{S_2}-X_{S_1})^2- (X_{M}-X_{S_1})^2  ] } 
\\
&& = \mu^2 \int_1^{+\infty} \frac{dv}{v} \int_v^{+\infty}  \frac{dw}{w}
\left[ 2 p_2(v,w) p_3(v,w)
 -4   p_1(v) p_2(v,w)  \right]
 \end{eqnarray}


The configurations of type $S_2'S_1M$ with the measure (\ref{tS2pS1M}) give
\begin{eqnarray}
&& \left[\Delta(\mu)\right]^{(2)}_{S_2'S_1M}
 \equiv  \left[ \Delta(\mu) \right]^{(1)+(2)}_{S_2'S_1M}
- \left[ \Delta(\mu) \right]^{(1)+(2)}_{S_1M}  
\\ &&  = 
\overline{   p_1(t;\tau_{S_2}) p_2(t;\tau_{S_2},\tau_{S_1})
   (X_{S_1} - X_{S_2'})^2
+  p_1(t;\tau_{S_2},\tau_{S_1}) p_3(t;\tau_{S_2},\tau_{S_1})  (X_M-X_{S_2'})^2 }   \nonumber \\ && + \overline{ [ p_2(t;\tau_{S_2},\tau_{S_1}) p_3(t\tau_{S_2},\tau_{S_1})
-  p_2(t;0,\tau_{S_1}) p_3(t;0,\tau_{S_1})  ](X_M-X_{S_1})^2 } 
 \\
&& = \mu^2 \int_1^{+\infty}  \frac{dv}{v} \int_1^{v} \frac{dw}{w} 
[2 p_1(v) p_2(v,w)
+ 6 p_1(v) p_3(v,w)  +  2 p_2(v,w) p_3(v,w)
-  2 p_1(w) (1-p_1(w)) ]
 \end{eqnarray}



The configurations of type $S_1ML_1$ with the measure (\ref{tS1ML1}) give
\begin{eqnarray}
&& \left[\Delta(\mu) \right]^{(2)}_{S_1ML_1}
 \equiv   \left[\Delta(\mu) \right]^{(1)+(2)}_{S_1ML_1}
- \left[\Delta(\mu) \right]^{(1)+(2)}_{S_1M}
- \left[\Delta(\mu) \right]^{(1)+(2)}_{ML_1}  
\\ &&  = 
\overline{   p_1(t;\tau_{S_1}) p_3(t;\tau_{S_1},\tau_M)
    [(X_{L_1} - X_{S_1})^2 -  (X_M - X_{S_1})^2] } \nonumber \\ && + \overline{ [ p_2(t;\tau_{S_1},\tau_M) p_3(t;\tau_{S_1},\tau_M)
- p_2(t;0,\tau_M) p_3(t;0,\tau_M)  ] (X_{L_1}-X_M)^2    } 
\\ 
&& = \mu^2 \int_1^{+\infty}  \frac{dv}{v} \int_0^1 \frac{dw}{w} 
\left[ 2  p_2(v,w) p_3(v,w)
+4  p_1(v) p_3(v,w)    
- 2 p_1(w) (1-p_1(w))   \right] 
 \end{eqnarray}

Finally, the sum of all contributions at order $\mu^2$ reads
\begin{eqnarray}
\left[ \Delta(\mu) \right]^{(2)}_{total}
&& \equiv   
\left[ \Delta(\mu) \right]^{(2)}_{S_2'S_1M} 
+  \left[ \Delta(\mu) \right]^{(2)}_{S_1M}
+  \left[ \Delta(\mu) \right]^{(2)}_{S_1S_2M} \nonumber \\ && + \left[ \Delta(\mu) \right]^{(2)}_{S_1ML_1} +  
\left[\Delta(\mu)\right]^{(2)}_{MI_2L_1}
+\left[ \Delta(\mu) \right]^{(2)}_{ML_1L_2} 
 + \left[ \Delta(\mu)  \right]^{(2)}_{ML_1} \nonumber \\
&& = - 4 \mu^2 \int_0^{+\infty} \frac{dv}{v} \ln v
  e^{-v}(1-e^{-v})  \\ 
&& + \mu^2 \int_0^{+\infty} \frac{dv}{v} \int_v^{+\infty}  \frac{dw}{w}
\left[ 4  p_2(v,w) p_3(v,w)  
-  6 p_1(v) p_2(v,w) +2 p_1(w) p_3(v,w) \right] +O(\mu^3)
 \end{eqnarray}
The double integral may be computed as in (\ref{double})
and yields
\begin{eqnarray}
\int_0^{+\infty} \frac{dv}{v} \int_v^{+\infty}  \frac{dw}{w}
\left[ 4  p_2(v,w) p_3(v,w)  
-  6 p_1(v) p_2(v,w) +2 p_1(w) p_3(v,w) \right]
= -\frac{\pi^2}{6} - 4 \ln 2
\end{eqnarray}
and thus the final result
\begin{eqnarray}
\left[ \Delta(\mu) \right]^{(2)}_{total}
=  \mu^2 [ 2 \ln 2 (\ln2+2 \gamma_E) -\frac{\pi^2}{6} - 4 \ln 2  ]  
 \end{eqnarray}
 coincides with the expansion of the exact result (\ref{widthexact}).

\subsection { Localization parameters at order $\mu^2$ }

\subsubsection{ Contributions at order $\mu^2$ of the two-traps configurations }

We have already studied the contributions of two traps configurations
when studying the order $\mu$.
The contribution of order $\mu^2$ of the configurations $ML_1$ (\ref{correcYkL1})
reads
\begin{eqnarray}
\overline{ [Y_k]^{(2)}_{ML_1} }
 = \mu^2 \int_0^{1} \frac{dv}{v} \ln v 
\left[ e^{- k v }   + (1-e^{-  v } )^k -1 \right]
\label{correcYkL1mu2}
\end{eqnarray}
whereas the contribution of order $\mu^2$ of the configurations $S_1M$ (\ref{correcYkS1})
reads
\begin{eqnarray}
\overline{ [Y_k]^{(2)}_{S_1M} }
 = - \mu^2 \int_1^{+\infty} \frac{dv}{v} \ln v  
[ e^{- k v  }
+  (1-e^{-  v } )^k -1 ]
\label{correcYkS1mu2}
\end{eqnarray}

\subsubsection{ Contributions at order $\mu^2$ of
 the three-traps configurations }

For a given configuration of three traps
 with occupation probabilities (\ref{occ3}), the localization parameters read
in terms of the variables $v \equiv \frac{t}{\tau_1}$
and $w \equiv \frac{t}{\tau_2}$
\begin{eqnarray}
Y_k  = p_1^k(v) + p_2^k(v,w)  + p_3^k(v,w) 
\end{eqnarray}

Following the procedure described above for the diffusion front,
we obtain the specific contributions at order $\mu^2$
of the various configurations as follows.


The configurations of type $ML_1L_2$  with the measure (\ref{tML1L2}) give
\begin{eqnarray}
\overline{[Y_k]^{(2)}_{ML_1L_2} } && = \overline{[Y_k]^{(0)+(1)+(2)}_{ML_1L_2}
- [Y_k]^{(0)+(1)+(2)}_{ML_1} }  \nonumber \\
&& = \mu^2 \int_0^1 \frac{dv}{v} \int_0^1  \frac{dw}{w}
\left[ p_2^k(v,w)+p_3^k(v,w)-(1-p_1(v))^k \right]+O(\mu^3)
\end{eqnarray}

The configurations of type $MLI_2L_1$  with the measure (\ref{tMI1L1}) give
\begin{eqnarray}
\overline{[Y_k]^{(2)}_{MI_2L_1} } && =
\overline{[Y_k]^{(0)+(1)+(2)}_{MI_2L_1}- [Y_k]^{(0)+(1)+(2)}_{ML_1} } \\
&& =  \mu^2 \int_0^1 \frac{dv}{v} \int_1^{+\infty}  \frac{dw}{w}
\left[ p_2^k(v,w)+p_3^k(v,w)-(1-p_1(v))^k \right]+O(\mu^3) 
\end{eqnarray}

The configurations of type $S_1S_2M$  with the measure (\ref{tS1S2M}) give
\begin{eqnarray}
\overline{[Y_k]^{(2)}_{S_1S_2M} } && =
\overline{[Y_k]^{(0)+(1)+(2)}_{S_1S_2M}- [Y_k]^{(0)+(1)+(2)}_{S_1M} } 
\nonumber \\
&&
 = \mu^2 \int_1^{+\infty} \frac{dv}{v} \int_v^{+\infty}  \frac{dw}{w} 
\left[ p_2^k(v,w)+p_3^k(v,w)-(1-p_1(v))^k \right]+O(\mu^3)
\end{eqnarray}

The configurations of type $S_2'S_1M$  with the measure (\ref{tS2pS1M}) give
\begin{eqnarray}
\overline{[Y_k]^{(2)}_{S_2'S_1M} } && =
\overline{[Y_k]^{(0)+(1)+(2)}_{S_2'S_1M}- [Y_k]^{(0)+(1)+(2)}_{S_1M} }
\nonumber  \\
&& =  \mu^2 \int_1^{+\infty} \frac{dv}{v} \int_1^{v}  \frac{dw}{w} 
\theta(v>w) \left[ p_1^k(v)+p_2^k(v,w)+p_3^k(v,w)-p_1^k(w)-(1-p_1(w))^k \right]+O(\mu^3)
\end{eqnarray}

The configurations of type $S_1ML_1$  with the measure (\ref{tML1L2}) give
\begin{eqnarray}
&& \overline{ [Y_k]^{(2)}_{S_1ML_1} } \equiv 
\overline{ [Y_k]^{(0)+(1)+(2)}_{S_1ML_1} - [Y_k]^{(0)} - 
[Y_k]^{(1)+(2)}_{S_1M}- [Y_k]^{(1)+(2)}_{ML_1} } \\
&& = 
\mu^2 \int_1^{+\infty}  \frac{dv}{v} \int_0^1 \frac{dw}{w} 
[ p_2^k(v,w)+p_3^k(v,w)+1 - (1-p_1(v))^k
- p_1^k(w)-(1-p_1(w))^k ]
+O(\mu^3)
\end{eqnarray}

The sum of all contributions of order $\mu^2$ thus reads
\begin{eqnarray}
&& \overline{[Y_k]^{(2)}_{total} } =\overline{[Y_k]^{(2)}_{MI_2L_1} } + \overline{ [Y_k]^{(2)}_{S_1ML_1} }
+ \overline{[Y_k]^{(2)}_{ML_1L_2} } + \overline{ [Y_k]^{(2)}_{ML_1} }+\overline{[Y_k]^{(2)}_{S_1S_2M} } + \overline{[Y_k]^{(2)}_{S_2'S_1M} }
+ \overline{ [Y_k]^{(2)}_{S_1M} } \nonumber \\
&& =
  \mu^2 \int_0^{+\infty} \frac{dv}{v} \int_v^{+\infty}  \frac{dw}{w} 
[ p_2^k(v,w)+p_2^k(w,v)+2 p_3^k(v,w)+1-p_1^k(v) 
 -2 (1-p_1(v))^k
 -(1-p_1(w))^k ]
\end{eqnarray}

For the special case $k=2$, we find
\begin{eqnarray}
\overline{[Y_2]^{(2)}_{total} } = \mu^2 \left[ 4 \ln 2 - \frac{\pi^2}{6} \right]
\end{eqnarray}
in agreement with the expansion of the exact result (\ref{y2exactexpansion}).
For the special case $k=4$, we find
\begin{eqnarray}
\overline{[Y_4]^{(2)}_{total} }= \mu^2 \left( 
- \frac{\pi^2}{6} + 2(21 \ln 2 + \ln 3 ( \ln 3 -12 )) \right)
\end{eqnarray}

\subsection { Correlation function at order $\mu^2$ }

\subsubsection{ Contributions at order $\mu^2$ of the two-traps configurations }

We have already studied the contributions of two traps configurations
when studying the order $\mu$.
The contribution of order $\mu^2$ of the configurations $ML_1$ (\ref{correML1})
reads
\begin{eqnarray}
\left[ {\cal C}_{\mu}(\lambda) \right]^{(2)}_{ML_1} 
= e^{- \lambda } 
 2 \mu^2 \int_0^1 \frac{dv}{v}  \ln v
 e^{-  v }   (1-e^{-  v } ) 
\end{eqnarray}

whereas the contribution of order $\mu^2$ of the configurations $S_1M$ (\ref{correS1M})
reads
\begin{eqnarray}
\left[ {\cal C}_{\mu}(\lambda) \right]^{(2)}_{S_1M}
= - 2 \mu^2 \lambda e^{-\lambda} \int_1^{+\infty} \frac{dv}{v} \ln v
 e^{-  v }   (1-e^{-  v } ) 
 \end{eqnarray}

\subsubsection{ Contributions at order $\mu^2$
 of the three-traps configurations }

For a given configuration of three traps situated at $(n_1,n_2,n_3)$
 with occupation probabilities $(p_1,p_2,p_3)$, the 
two-particles correlation function reads
\begin{eqnarray}
C(l,t) = (p_1^2+p_2^2+p_3^2) \delta_{l,0}
+ 2 p_1 p_2 \delta_{l,n_2-n_1} 
 + 2 p_1 p_3 \delta_{l,n_3-n_1}
 + 2 p_2 p_3 \delta_{l,n_3-n_2}
\end{eqnarray}

Since the weight of the delta pic is given by the localization 
parameter $Y_2$ that we have already considered above,
we will consider in the following
only the scaling function ${\cal C}_{\mu}(\lambda)$ (\ref{correform}).
Following the procedure described above for the diffusion front,
we will obtain the specific contributions at order $\mu^2$.


The configurations of type $ML_1L_2$ give
the specific contribution at order $\mu^2$
\begin{eqnarray}
&& \left[ {\cal C}_{\mu}(\lambda) \right]^{(2)}_{ML_1L_2}
 = \left[ {\cal C}_{\mu}(\lambda) \right]^{(1)+(2)}_{ML_1L_2}
- \left[ {\cal C}_{\mu}(\lambda) \right]^{(1)+(2)}_{ML_1}
\nonumber \\
&& = e^{-\lambda} 
\mu^2 \int_0^1 \frac{dv}{v} \int_0^1  \frac{dw}{w}
\left[     2 p_2(v,w) p_3(v,w) -2 p_1(v) p_3(v,w)
 + \lambda 2 p_1(v) p_3 (v,w) 
 \right]  +O(\mu^3)
 \end{eqnarray}


The configurations of type $MI_2L_1$ give
the specific contribution at order $\mu^2$
\begin{eqnarray}
&&  \left[ {\cal C}_{\mu}(\lambda) \right]^{(2)}_{MI_2L_1}
 = \left[ {\cal C}_{\mu}(\lambda) \right]^{(1)+(2)}_{MI_2L_1}
- \left[ {\cal C}_{\mu}(\lambda) \right]^{(1)+(2)}_{ML_1}
\nonumber \\
&& = e^{-\lambda} 
\mu^2 \int_0^1 \frac{dv}{v} \int_1^{+\infty}  \frac{dw}{w}
\left[ 2 p_1(v) p_2(v,w)   + 2 p_2(v,w) p_3(v,w) 
 - 2 p_1(v) p_2 (v,w)\lambda  \right]  +O(\mu^3)
\end{eqnarray}


The configurations of type $S_1S_2M$ give
the specific contribution at order $\mu^2$
\begin{eqnarray}
&&  \left[ {\cal C}_{\mu}(\lambda) \right]^{(2)}_{S_1S_2M}
 = \left[ {\cal C}_{\mu}(\lambda) \right]^{(1)+(2)}_{S_1S_2M}
- \left[ {\cal C}_{\mu}(\lambda) \right]^{(1)+(2)}_{S_1M}
\nonumber \\
&& = e^{-\lambda} 
\mu^2 \int_1^{+\infty} \frac{dv}{v} \int_v^{+\infty}  \frac{dw}{w}
\left[ 2 p_1(v) p_2(v,w)   + 2 p_2(v,w) p_3(v,w) 
 - 2 p_1(v) p_2 (v,w)\lambda  \right]  +O(\mu^3)
 \end{eqnarray}


The configurations of type $S_2'S_1M$ give
the specific contribution at order $\mu^2$
\begin{eqnarray}
&&  \left[ {\cal C}_{\mu}(\lambda) \right]^{(2)}_{S_2'S_1M}
 = \left[ {\cal C}_{\mu}(\lambda) \right]^{(1)+(2)}_{S_2'S_1M}
- \left[ {\cal C}_{\mu}(\lambda) \right]^{(1)+(2)}_{S_1M}
\nonumber \\
&& = e^{-\lambda} 
\mu^2 \int_1^{+\infty} \frac{dv}{v} \int_v^{+\infty}  \frac{dw}{w}
  \left[ 2 p_1(w) p_2(w,v)   + 2 p_2(w,v) p_3(v,w)  -2 p_1(v) (1-p_1(v))
 + 2 p_1(w) p_3 (v,w)\lambda
   \right]  +O(\mu^3)
 \end{eqnarray}


The configurations of type $S_1ML_1$ give
the specific contribution at order $\mu^2$
\begin{eqnarray}
&& \left[ {\cal C}_{\mu}(\lambda) \right]^{(2)}_{S_1ML_1}
 = \left[ {\cal C}_{\mu}(\lambda) \right]^{(1)+(2)}_{S_1ML_1}
- \left[ {\cal C}_{\mu}(\lambda) \right]^{(1)+(2)}_{S_1M}
 - \left[ {\cal C}_{\mu}(\lambda) \right]^{(1)+(2)}_{ML_1} 
\nonumber \\
= && \mu^2 e^{-\lambda} \int_0^{1}  \frac{dv}{v} \int_1^{+\infty} \frac{dw}{w}
 \left[ 2 p_2(w,v) p_3(v,w)-2 p_1(w) p_3(w,v)    -2 p_1(v) (1-p_1(v))
 + 2 p_1(w) p_3 (v,w) \lambda
   \right]  +O(\mu^3)
 \end{eqnarray}


The sum of all contributions of order $\mu^2$ reads
\begin{eqnarray}
 \left[ {\cal C}_{\mu}(\lambda) \right]^{(2)}_{total}
&& = \left[ {\cal C}_{\mu}(\lambda) \right]^{(2)}_{MI_2L_1}
+  \left[ {\cal C}_{\mu}(\lambda) \right]^{(2)}_{S_1ML_1}
+ \left[ {\cal C}_{\mu}(\lambda) \right]^{(2)}_{S_1S_2M}
+  \left[ {\cal C}_{\mu}(\lambda) \right]^{(2)}_{S_2'S_1M}
+\left[ {\cal C}_{\mu}(\lambda) \right]^{(2)}_{S_1M} 
\nonumber  \\
&&+ \left[ {\cal C}_{\mu}(\lambda) \right]^{(2)}_{ML_1L_2}
+ \left[ {\cal C}_{\mu}(\lambda) \right]^{(2)}_{ML_1}
\nonumber  \\
&& = \mu^2 e^{- \lambda} \left[  \frac{\pi^2}{3}-\ln 2(4+\ln 2 +\gamma_E)
+\lambda \left( -\frac{\pi^2}{6}+\ln 2 (\ln 2 +\gamma_E) \right) \right]
 \end{eqnarray}

\section {  Hierarchical structure of the important traps}

\label{hierarchie}

It is now clear that the procedure that we have described
up to order $\mu^2$ can be generalized 
at an arbitrary order $n$ : all observables at order $\mu^n$
can be obtained by considering a dispersion of the thermal packet
over at most $(1+n)$ traps, that have to be chosen among a 
certain number
$\Omega_n$ of possible configurations of the traps. 
Our aim in this Section is not to pursue any further explicit computations
but to get some insight into the set of important
traps that play a role at a given order $n$.

\subsection{Set of the important traps at order $n$}

At order $n$, the important traps are :

\begin{itemize}

\item the main trap $M$

\item the following $n$ large renormalized traps $L_1$,...$L_n$

\item the $n$ biggest traps $S_1$...$S_n$ among the small traps before $M$

\item the $(n-1)$ biggest traps $I_2^{(1)}$...$I_n^{(1)}$ among the small traps
in the interval between $M$ and $L_1$

\item the $(n-2)$ biggest traps $I^{(2)}_3$...$I^{(2)}_n$ among the small traps in the interval
between $L_1$ and $L_2$

\item...

\item the biggest trap 
$I^{(n-1)}_n$ among the small traps in the interval
between $L_{n-2}$ and $L_{n-1}$. 

\end{itemize}

The index at the bottom
represents the order of occupation in $\mu$ as in the Figure
\ref{trap}.  
The total number of traps is thus  
\begin{eqnarray} 
T_n=1+n+\sum_{i=1}^n i = 1+\frac{n (n+3)}{2}
\end{eqnarray}
which generalizes $T_1=3$ ($M$,$S_1$,$L_1$)
and $T_2=6$ ($M$,$S_1$,$S_2$ identified with $S_2'$,$L_1$,$L_2$).  

\subsection{Set of the important configurations at order $n$}

With these $T_n$ traps, we have now to construct the possible 
$\Omega_n$ configurations of
$(1+n)$ traps, that are ordered by in positions,
and that contribute
up to order $\mu^n$. 
We have
\begin{eqnarray} 
\Omega_n=\Omega_{n-1}+\omega_n = \sum_{i=0}^n \omega_i
\end{eqnarray}
 where $\omega_n$
represents the number of configurations that begin to contribute
at order $n$. 

We may now decompose
\begin{eqnarray} 
\omega_n=a_n^{(L_n)}+a_n^{(L_{n-1})}+...a_n^{(L_1)}+a_n^{M}
\end{eqnarray}
where $a_n^{(L_j)}$ is the number of configurations
that contain $L_j$ as the rightmost trap.
For $j=n$, there is only $a_n^{(L_n)}=1$ configuration $ML_1L_2...L_n$,
whereas for $j=0$
\begin{eqnarray} 
a_n^{M}=n!
\end{eqnarray}
since we have to order in space the $n$ traps $S_1$...$S_n$ before $M$.
More generally, at order $j$, to construct the configurations 
of $(n+1)$ traps containing $M L_1...L_{j}$,
which represent $(j+1)$ fixed traps,
we have to choose $(n-j)$ traps among the $(j+1)$ available intervals
and to count the possible positional orders in each interval
\begin{eqnarray} 
a_n^{L_j}=\sum_{p_1=0}^{+\infty} ... \sum_{p_{j+1}=0}^{+\infty}
\delta(\sum_{i=1}^{j+1} p_i=n-j) p_1! ... p_{j+1}!
\end{eqnarray}

The final result is thus that the number of new configurations
that appear at order $n$ reads
 \begin{eqnarray} 
\omega_n=\sum_{j=0}^n \left[\sum_{p_1=0}^{+\infty} ... \sum_{p_{j+1}=0}^{+\infty}
\delta(\sum_{i=1}^{j+1} p_i=n-j) p_1! ... p_{j+1}!
\right]
\end{eqnarray}
which generalize what we have found before for the lowest orders
$\omega_0=1$ ($M$),
$\omega_1=2$ ($S_1M$ and $ML_1$) and $\omega_2=5$ ($S_1S_2M$, $S_2'S_1M$, $MI_1L_1$, $ML_1L_2$ and $S_1ML_1$).

\section{ Quantitative mapping between the biased Sinai diffusion 
and the directed trap model }

\label{sinaibias}

\subsection{ Renormalized landscape for the biased Brownian motion }

The real space renormalization group (RSRG) method
can also be applied to the biased Brownian landscape 
\cite{daniel,us_sinai}. 
The distributions of the barriers $F=\Gamma+\xi$ 
in the renormalized landscape at scale $\Gamma$ are given by
\cite{daniel,us_sinai,us_toy}
\begin{eqnarray} 
P^+_{\Gamma}(\xi) && = \frac{2 \delta }{e^{2 \delta \Gamma}-1} e^{-\xi \frac{2 \delta }{e^{2 \delta \Gamma}-1}} \simeq  \frac{2 \delta }{e^{2 \delta \Gamma}} e^{-\xi \frac{2 \delta }{e^{2 \delta \Gamma}}} \\
P^-_{\Gamma}(\xi) && = \frac{2 \delta }{1-e^{-2 \delta \Gamma}}
 e^{-\xi \frac{2 \delta }{1-e^{-2 \delta \Gamma}}}
\simeq 2 \delta 
 e^{-\xi 2 \delta }
\label{rgdrift}
\end{eqnarray}
where the parameter $2\delta $ reads in terms of the notations (\ref{langevin},
\ref{defmu}) 
\begin{eqnarray}
2 \delta \equiv  \frac{\mu}{T}= \frac{F_0}{\sigma}
\end{eqnarray}

As a consequence, the distribution $P^-_{\Gamma}(\xi)$
of barriers against the bias
can be considered as infinitely large only in the limit of vanishing bias
$\delta \to 0$. It is only in this limit
that all particles of the same thermal packet
remain in the same renormalized valley asymptotically.

\subsection{Trapping-time of a renormalized valley of barrier $\Gamma$} 

Let us now recall a standard result for one-dimensional Fokker-Planck equation
\cite{gardiner} : for a particle diffusing in a potential $U(x)$ on a interval $[a,b]$ with reflecting condition at $a$, 
the exit time defined as the first-passage time $\theta(x)$
 at the point $b$ for a particle starting at $x \in (a,b)$ at time $t=0$
can be studied for an arbitrary potential $U(x)$ :
the moments   
\begin{eqnarray}
\theta_n(x) \equiv < [\theta(x)]^n > 
\end{eqnarray}
are given by the recurrence
\begin{eqnarray}
\theta_n(x) = \beta \int_x^b dy e^{\beta U(y)} \int_a^{y} dz e^{-\beta U(z)}
[ n \theta_{n-1}(z)]
\label{recurrencetheta}
\end{eqnarray}
with the initial condition $\theta_{n=0}(x)=1$.
In particular, the first moment reads
\begin{eqnarray}
\theta_1(x) = \beta \int_x^b dy e^{\beta U(y)} \int_a^{y} dz e^{-\beta U(z)}
\label{theta1ab}
\end{eqnarray}

\begin{figure}[thb]

\centerline{\fig{8cm}{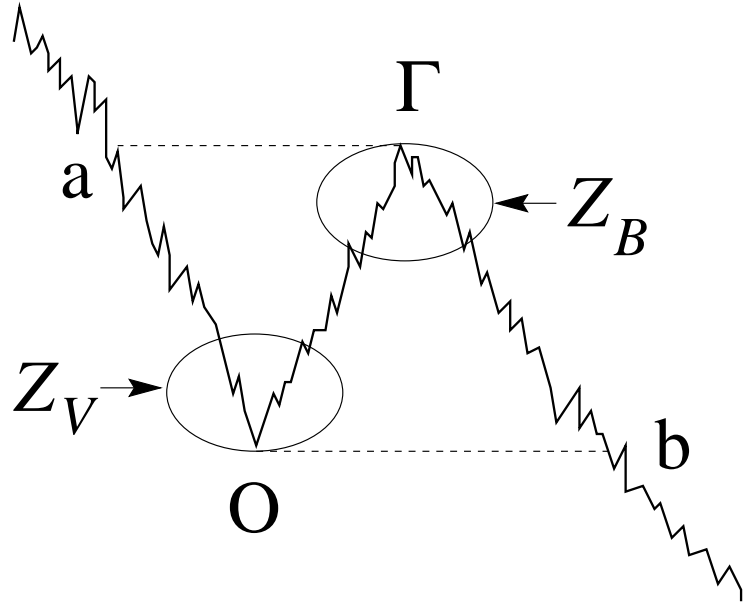}} 
\caption{ Computation of the escape time from a renormalized valley
of barrier $\Gamma$ : we consider the first-passage-time at $b$
for a particle starting at $0$. The double integral (\ref{theta1ab})
is dominated by the Arrhenius factor $e^{\beta \Gamma}$, and the prefactor
is the product of two partition function :
$Z_V$ represents the partition function
of the bottom of the valley and $Z_B$ represents the partition function
of the inverse potential $(-V)$ near the top of the barrier $\Gamma$.}
 \label{escapeti}  

\end{figure}

For the biased Brownian landscape, the exit time over a barrier $\Gamma$
when starting at the bottom of a renormalized valley
that we choose as the origin   
can be obtained (see Figure \ref{escapeti})
by choosing $a$ at the height $\Gamma$ on the renormalized descending bond
on the left and $b$ at a potential $\Gamma$ after the top
of the barrier $\Gamma$.  
It seems that usually \cite{gardiner}
one chooses $b$ exactly at the top of the barrier to derive the
Arrhenius factor, but we think that to obtain the correct prefactor, 
one has to choose $b$ on the descending potential {\it after}
the top to be sure that the particle
will not return in the trap where it started.
Indeed, when the particle sits just on the top, there is a finite probability
to return to its starting trap, which is for instance a probability
$1/2$ for a potential that is symmetric around its top. 
So for a given realization $V$ of a renormalized valley, 
the first moment of the escape time reads
\begin{eqnarray}
\theta_1\{V\} = \beta \int_0^{b} dy e^{\beta V(y)} 
\left[ \int_a^{0} dz e^{-\beta V(z)}
+ \int_0^{y} dz e^{-\beta V(z)} \right]
\label{theta1}
\end{eqnarray}
where the biased Brownian potential $
V(x)  = - F_0 x + U(x) $
satisfies the constraints of a renormalized valley
at scale $\Gamma$ (see Figure \ref{escapeti}) :
it starts at $V(0)=\epsilon$, 
it then evolves on each side $x>0$ and $x<0$ 
in the presence of absorbing boundaries
at $0$ and $\Gamma$, and is conditioned to 
finish at $V=\Gamma$ and not at $V=0$.  
On the negative side, $(x=a)$ is the random position
where the potential first hit $\Gamma$.
On the positive side, after the random position  
 $l_{\Gamma}$ where the potential first hit $\Gamma$, 
the potential again evolve in the presence of absorbing boundaries
at $0$ and $\Gamma$, and is conditioned to 
finish at $V=0$, at some random position called $b$,
and not at $V=\Gamma$.

As usual for the problem of escape over a large barrier, 
the double integral (\ref{theta1}) is dominated by the saddle 
where $V(y)$ is maximal and where $V(z)$ is minimal.
For a renormalized valley (see Figure \ref{escapeti}), 
these regions are $y \sim l_{\Gamma}^{(1)}$
where $V(y) \sim \Gamma$ and $z \sim 0$ where $V(z) \sim 0$.
This saddle-point analysis yields the following leading behavior
\begin{eqnarray}
\theta_1 \{V \} \opsimeq_{\Gamma \to \infty} \tau_0(V) e^{\beta \Gamma}
\label{theta1V}
\end{eqnarray}
The prefactor is simply given by the product
\begin{eqnarray}
 \tau_0(V) = \beta Z_B Z_V
\label{tauoV}
\end{eqnarray}
where $Z_V$
is the partition function of the infinitely deep renormalized valley  
\begin{eqnarray}
Z_V = \lim_{\Gamma \to \infty} 
\left( \int_0^{l_{\Gamma}^{(2)}} dz e^{-\beta V_-(z)}
+ \int_0^{l_{\Gamma}^{(1)}} dz e^{-\beta V_+(z)} \right)
\label{zvdef}
\end{eqnarray}
where the random potentials 
\begin{eqnarray}
V_+(x) && = - F_0 x + U_1(x)  \\
V_-(x) && = F_0 x + U_2(x) 
\label{vpm}
\end{eqnarray}
are defined in terms of two
independent Brownian trajectories $U_1(x)$ and $U_2(x)$ 
(\ref{defsigma}) starting at $V_+(0)=\epsilon=V_-(0)$. 
The potentials $V_{\pm}(x)$ evolves in the presence of absorbing boundaries
at $0$ and $\Gamma$, and are conditioned to 
finish at $V=\Gamma$ and not at $V=0$. 
 $l_{\Gamma}^{(1)}$ and $l_{\Gamma}^{(2)}$ are the random times where 
$V_{\pm}$ respectively first hit $V=\Gamma$.

Similarly, the factor $Z_B$ 
is the partition function of an independent 
infinitely deep renormalized valley,
 which represents what happens in the vicinity
of the top of the barrier $\Gamma$ when considered
with the change $V \to -V$ to transform it in a valley
(see Figure \ref{escapeti}).
For the biased Brownian landscape considered here, by symmetry, 
$Z_B$ is simply an independent realization of the variable $Z_V$.

The same saddle point analysis may be applied to
higher moments given by the recurrence (\ref{recurrencetheta})
to obtain 
\begin{eqnarray}
\theta_n \{V \} \opsimeq_{\Gamma \to \infty}
n! ( \tau_0(V) e^{\beta \Gamma} )^n
\end{eqnarray}

So for a given renormalized valley of barrier $\Gamma$, the escape time $t$
is distributed exponentially as in the trap model
\begin{eqnarray}
f_{\theta_1 \{V\} } (t) = \frac{1}{\theta_1 \{V\}} e^{- \frac{t}{\theta_1 \{V\}}}
\end{eqnarray}
where the trapping-time $\theta_1 \{V\}$ (\ref{theta1V})
depends mostly on the barrier $\Gamma$ via the usual Arrhenius factor $e^{\beta \Gamma}$, but also on the details of the structure of the 
valley near the bottom and near the top via the prefactor (\ref{tauoV}).

\subsection{Distribution of the trapping-time of renormalized valleys} 

We are now interested into the distribution of the trapping-time 
$\theta_1 \{V\}$ over the ensemble of renormalized valleys
existing in the renormalized landscape at scale $\Gamma$.
The distribution of the barriers is given by (\ref{rgdrift}).
So we have to study the statistics of the prefactor (\ref{tauoV}).

It is more convenient to work with dimensionless quantities
by rewriting the partition functions as $Z_V=z_1/(\sigma \beta^2)$,
$Z_B=z_2/(\sigma \beta^2)$
so that
\begin{eqnarray}
 \tau_0(V) = \frac{e^{\beta \Gamma} }{\sigma^2 \beta^3} z_1 z_2
\label{tauoVbis}
\end{eqnarray}
where $z_1$ and $z_2$ are independent random variables
whose probability distribution $P(z)$ is characterized in Appendix 
\ref{pathbias} by its Laplace transform 
\begin{eqnarray}
\int_0^{+\infty} dz e^{- s z} P(z) && = \left[ \frac{1}{\Gamma(1+\mu)}
\frac{(\sqrt s)^{\mu}  }{I_{\mu} (2 \sqrt s )} \right]^2 
\label{laplacez}
\end{eqnarray}
where $I_{\mu}$ is the Bessel function of index $\mu$.
In the renormalized landscape at scale $\Gamma$, the probability distribution
of the trapping time $\tau$ of the renormalized valleys thus reads
using (\ref{rgdrift})
\begin{eqnarray}
P_{\Gamma}(\tau) && = \int_0^{+\infty} d\xi 2\delta e^{-2 \delta \xi}
\int_0^{+\infty} dz_1  P(z_1) \int_0^{+\infty} dz_2 P(z_2)
\ \delta \left[ \tau-  \frac{e^{\beta (\Gamma+\xi)} }{\sigma^2 \beta^3} z_1 z_2 \right] \\
&& = \frac{\mu}{\tau} 
\left( \frac{e^{\beta \Gamma} }{ \sigma^2 \beta^3 \tau} \right)^{\mu}
\int_0^{+\infty} dz_1 z_1^{\mu} P(z_1) \int_0^{+\infty} dz_2 (z_2)^{\mu} P(z_2)
\label{ptau}
\end{eqnarray}
So we have to compute the non-integer moment of order $\mu$ of the variable
$z$. Using the integral representation valid for $0<\mu<1$
\begin{eqnarray}
 z^{\mu} = \frac{\mu}{\Gamma(1-\mu) } \int_0^{+\infty} ds s^{-1-\mu}
\left(1- e^{-s z} \right)
\end{eqnarray}
we obtain the moment from the Laplace transform (\ref{laplacez})
\begin{eqnarray}
\int_0^{+\infty} dz z^{\mu} P(z)
&& = \frac{\mu}{\Gamma(1-\mu) } \int_0^{+\infty} ds s^{-1-\mu}
  \left(1- \left[ \frac{1}{\Gamma(1+\mu)}
\frac{(\sqrt s)^{\mu}  }{I_{\mu} (2 \sqrt s )} \right]^2 \right) \\
&& = \frac{2 \mu 4^{\mu}}{\Gamma(1-\mu) } 
 \int_0^{+\infty} dw 
\left[  w^{-1-2\mu} - \frac{ 1}{\Gamma^2(1+\mu) 4^{\mu} w I_{\mu}^2(w)}
\right]
\label{momentz}
\end{eqnarray}
Using the wronskian property of Bessel functions, and their series expansion at small argument, we finally get
\begin{eqnarray}
\int_0^{+\infty} dz z^{\mu} P(z)
&& =  
 \frac{1}{\Gamma(1-\mu) } 
 \lim_{a \to 0}
\left[ \left( \frac{a}{2} \right)^{-2 \mu} - 
\frac{2 \mu}{ \Gamma^2(1+\mu)}  
\frac{K_{\mu}(a)}{   I_{\mu}(a)}  \right]
  =  \frac{1}{\Gamma(1+\mu) } 
\label{momentzbis}
\end{eqnarray}
The final result is thus that the distribution of trapping time $\tau$
of the renormalized valleys existing
at scale $\Gamma$ reads (\ref{ptau})
\begin{eqnarray}
P_{\Gamma}(\tau) 
&& = \frac{\mu}{\tau} 
\left( \frac{e^{\beta \Gamma} }{ \sigma^2 \beta^3 \tau} \right)^{\mu}
\frac{1}{\Gamma^2(1+\mu) } 
\label{ptaug}
\end{eqnarray}

\subsection{Precise choice of the renormalization scale $\Gamma$ as a function of time}

We have seen in the trap model that the distribution
of renormalized traps at $t$ reads (\ref{tauM})
\begin{eqnarray}
q_t(\tau)= \theta(t<\tau) \frac{\mu}{\tau} 
\left( \frac{t}{\tau} \right)^{\mu}
\label{qttau}
\end{eqnarray}
To make it exactly coincide with the distribution (\ref{ptaug})
of the biased Sinai model, we have to choose the renormalized scale $\Gamma$
of the landscape to be the following function of time
\begin{eqnarray}
\Gamma(t) = T \ln \left[ t \sigma^2 \beta^3 
\left( \Gamma^{2} (1+\mu) \right)^{\frac{1}{\mu}}  \right]
\label{gammat}
\end{eqnarray}

The RSRG method \cite{us_sinai} gives that
the distribution of the length $l_+$ of the descending renormalized bonds
is simply exponential in the limit $ \Gamma \to \infty$
\begin{eqnarray}
P_{\Gamma}(l_+) =
 \frac{1}{b_{\Gamma}} e^{- \frac{l_+}{b_{\Gamma}} }
\label{plplus}
\end{eqnarray}
where the mean length reads
\begin{eqnarray}
b_{\Gamma}=  \frac{1}{\sigma (2 \delta)^2} e^{2 \delta \Gamma } 
= \frac{ 1 }{\sigma \beta^2 \mu^2} e^{2 \delta \Gamma }
\label{bgamma}
\end{eqnarray}
so that it reads as a function of time (\ref{gammat})
\begin{eqnarray}
b(t)= b_{\Gamma(t)}  
= \frac{ \Gamma^2 (1+\mu) }{\sigma \beta^2 \mu^2} 
\left[ t \sigma^2 \beta^3   \right]^{\mu}
= \frac{ \Gamma^2 (\mu) }{\sigma \beta^2 } 
\left[ t \sigma^2 \beta^3   \right]^{\mu}
\label{bt}
\end{eqnarray}
This length scale $b(t)$ exactly corresponds to the ratio $t^{\mu} 
\frac{c_{trap}(\mu)}{c_{sinai}(\mu)}$
of the constants 
appearing in the exact diffusion front of the two models (\ref{ratio}).

\subsection{Usual RSRG in the limit $\mu \to 0$} 

It has been shown in \cite{us_sinai} that the ``effective dynamics",
where at time $t$,
the particle  
is typically at time $t$ around the minimum of the renormalized valley
containing the initial condition,
is sufficient in the double limit $t \to \infty$ $\delta \to 0$
with 
\begin{eqnarray}
\gamma \equiv  \delta \Gamma(t)
\end{eqnarray}
fixed and $X= \frac{x}{\Gamma^2(t)}$ fixed.

The limit $\gamma \to 0$ corresponds to the symmetric Sinai diffusion,
whereas in the limit
  $\gamma \to \infty$, the model becomes directed at large scale
and the diffusion front converges towards \cite{us_sinai}
\begin{eqnarray}
\overline{P(x,t \vert 0,0)} \opsimeq_{\gamma \to \infty}
\theta(x) \frac{1}{b(t)} e^{- \frac{x}{b(t)} }
\label{lowestfront}
\end{eqnarray}
where $b(t)$ represents the mean length of renormalized descending bonds (\ref{bt}).
This limit actually corresponds to 
the limit $\mu \to 0$ of the exact Levy front 
\cite{annphys,tanaka,hushiyor} as described in the Appendix (\ref{levy}).

\subsection{ Spreading of the thermal packet over many renormalized valleys}

The renormalized valleys of the Sinai model with bias
are the analog of the traps in the directed model .
For $\mu \to 0$, the bottom of the renormalized valley
containing the origin described above
 (\ref{lowestfront}) is the analog of the main trap M
described in section \ref{maintrap}.

\begin{figure}[thb]

\centerline{\fig{8cm}{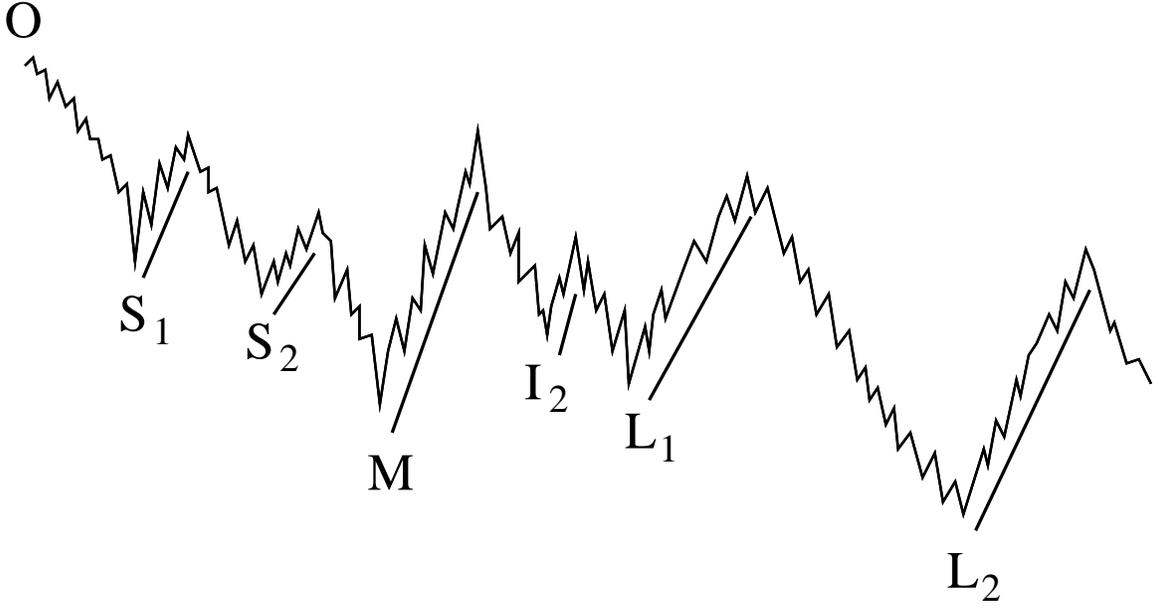}} 
\caption{ Hierarchical structure of the important valleys for a particle starting at the origin.
The barriers against the bias that are emphasized by the straight lines
correspond to the depths of the trap model represented on Figure \ref{trap}.  The bottom $M$ of the renormalized valley that contains the origin
at scale $\Gamma$ is occupied with
a weight of order $O(\mu^0)$. The bottom $L_1$ of the next renormalized
valley and the bottom $S_1$ of the biggest 
sub-valley before $M$ are occupied with weights of order $O(\mu)$.
The next-nearest renormalized valley $L_2$, 
the biggest sub-valley $I_2$ between $M$ and $L_1$,
and the second biggest sub-valley $S_2$ before $M$ are occupied
with weights of order $O(\mu^2)$. }
\label{sinai}  

\end{figure}

At first order in $\mu$, as in Section \ref{dispersionmu},
there are two effects :

\begin{itemize} 

\item Next renormalized valley $L_1$

The main renormalized valley $M$ at scale $\Gamma(t)$ has a trapping time
$\tau_M$ which is distributed as in the directed trap model (\ref{tauM}),
since we have defined the relation $\Gamma(t)$ (\ref{gammat})
by identifying the trapping time distributions of the two models.
So there is a small probability $(1-e^{-\frac{t}{\tau_M}})$ that the particle has already escaped from this main renormalized $M$
at time $t$, to jump into the next renormalized valley $L_1$.

Moreover, the RSRG approach \cite{us_sinai}
yields that the joint distribution
of the trapping time $\tau_M$  
and of the positions $x_M$ and $x_{L}$ of the bottoms
of the main renormalized valley and of the 
the next renormalized valley $L_1$ reads
\begin{eqnarray}
{\cal D}_{M,L_1}(x,x_{L};\tau_M) &&  
= \theta(t<\tau_M) \theta(0<x<x_L) \frac{\mu t^{\mu} }{\tau_M^{1+\mu} } \frac{1}{b^2(t)}
e^{- \frac{x}{b(t)} } e^{- \frac{(x_L-x)}{b(t)} } \\
&& = \theta(t<\tau_M) \theta(0<x<x_L) \frac{\mu t^{\mu}}{\tau_M^{1+\mu} } \frac{1}{b^2(t)}
e^{- \frac{x_L}{b(t)} } 
\end{eqnarray}
which is the analog of (\ref{premeasureL}).
The only change is in the prefactor in front of $t^{\mu}$
in the scale $b(t)$ (\ref{bt}).

\item Last decimated renormalized valley $S_1$

The last decimated barrier against the bias
inside the main renormalized
valley between the origin and the bottom,
defines the last decimated sub-valley $S_1$ : 
it has a trapping time $\tau_{S_1}<t$ which is not zero and thus there is a small probability $e^{-\frac{t}{\tau_{S_1}}}$ that the particle
 is still trapped in the sub-valley $S_1$ at time $t$.

Moreover, the RSRG approach \cite{us_sinai}
yields that the joint distribution
of the trapping time $\tau_{S}$  
and of the positions $x_{S}$ and $x$ of the bottoms
of the last decimated valley $S_1$ and of the 
the main renormalized valley $M$ reads
\begin{eqnarray}
{\cal D}_{S_1,M}(x_S,x;\tau_S) &&  
= \theta(t>\tau_S) \theta(0<x_S<x) \frac{\mu}{\tau_S} \frac{1}{b(t) b(\tau_S) }
e^{- \frac{x}{b(\tau_S) } }
\end{eqnarray}
which is the analog of (\ref{premeasureS}).
The only change is again in the prefactor 
in the scale $b(t)$ (\ref{bt}). 

\end{itemize}

It is clear that this analysis may be generalized
to further orders in $\mu$.

\subsection{ Conclusion : Equivalence of the two large-scale renormalized descriptions }

The statistical properties of the
spreading of the thermal packet over many renormalized valleys
and sub-valleys inside the main one are thus exactly the same as
in the directed trap model discussed in details in 
the first Sections.
In particular, the localization parameters $Y_k$
of the trap model represent coarse-grained localization
parameters for the biased Sinai diffusion : ``at the same position"
 in the trap model means ``at a finite distance around the 
bottom of the same renormalized valley" for the biased Sinai diffusion.
As a consequence, for all rescaled quantities $\frac{x}{t^{\mu}}$,
the results are exactly the same up to the global prefactor 
in the scale $b(t)$ (\ref{bt}) :
this was already known for the averaged diffusion front (see Appendix
\ref{levy}), but this also holds for the thermal width (\ref{widthsinai}),
for all other rescaled thermal cumulants (\ref{cumultrap}),
and for the long-range part of the two-point correlation function (\ref{corresinai}).

The new property of the Sinai model is thus the internal structure 
of a renormalized valley, that induces a dispersion
over finite distances
of the particles that are in the same renormalized valley.
We now study the statistical properties of the biased Brownian valleys.

\section{ Internal structure of the traps in the biased Sinai diffusion}

\label{internal}

\subsection{Probability distribution inside a renormalized valley}

The probability distribution of particles inside the same renormalized valley
can be obtained by generalizing the approach of the 
Sinai symmetric case \cite{us_golosov} : for each realization
of a renormalized valley, it is given by the Boltzmann distribution
on this valley. So asymptotically as $t \to \infty$, 
the probability distribution of the distance $y$ 
to the bottom of the valley, averaged over the environment reads
\begin{eqnarray} 
P_{V}(y >0) && =  \lim_{\Gamma \to \infty} \left< 
 \frac{e^{- \beta V_+( y )} }
{ \int_0^{l_{\Gamma}^{(1)}} dx e^{- \beta V_+(x)} 
+ \int_0^{l_{\Gamma}^{(2)}} dx e^{-\beta V_-(x)} }  \right>_{\{V_+\},\{V_-\}}
\nonumber \\
P_{V}(y <0) && =  \lim_{\Gamma \to \infty} \left< 
 \frac{e^{- \beta V_-( \vert y \vert)} }
{ \int_0^{l_{\Gamma}^{(1)}} dx e^{- \beta V_+(x)} 
+ \int_0^{l_{\Gamma}^{(2)}} dx e^{-\beta V_-(x)} }  \right>_{\{V_+\},\{V_-\}}
\label{boltzmannvalleybias}
\end{eqnarray}
where the random potentials $V_{\pm}$ satisfy the same conditions
as in (\ref{vpm}).

The computation of the functionals (\ref{boltzmannvalleybias})
is given in Appendix \ref{pathbias}.
It yields the non-intuitive result that the probability
distribution $P_V(y)$ is actually symmetric in $y \to -y$.
The restoration of this symmetry comes from the conditioning
of the biased random walk to reach $\Gamma$ on each side.
Its Laplace transform reads (\ref{pvpinter})
\begin{eqnarray}
{\hat P}_V(p)  \equiv \int_0^{+\infty} du e^{-p y} {\hat P}_V(y)  =
 = \frac{ 1}{ \Gamma^2(1+\mu )  }
\int_0^{\infty} ds
\frac{ \left(\frac{s}{2} \right)^{2 \mu-1} }{  I_{\mu}(s)
I_{\nu} (s) }
 \int_0^{s} dz  z  I_{\nu} (z)  \left[
K_{\mu}(z) 
- \frac{K_{\mu}(s)}
{ I_{\mu}(s)}
I_{\mu}(z)  \right]
\label{pvpfinal}
\end{eqnarray}
where the only factor containing the Laplace parameter is the index
\begin{eqnarray}
\nu \equiv \sqrt{\mu^2+ \frac{4 T^2 p}{\sigma}}=\mu +\frac{2 T^2 p}{\sigma \mu }+O(p^2)
\end{eqnarray}
For $\mu>0$, the series expansion in the Laplace parameter $p $ 
is thus regular
leading to
\begin{eqnarray}
{\hat P}_{V}(p=0) && = \frac{1}{2} - \frac{2 T^2 p}{\mu \sigma } D(\mu) 
+...
\end{eqnarray}
All moments are thus finite, contrary to the symmetric case $\mu=0$
\cite{us_golosov}, where the behavior as $(1-\sqrt{p} c+..)$
corresponds to the algebraic decay as $1/y^{3/2}$.
So in the biased case, the distribution inside a renormalized valley
is very narrow, contrary to the symmetric case. 

\subsection{Localization parameters inside a renormalized valley}

For $k$ particles that are in the same renormalized valley,
the localization parameters may be computed as
an average of the k-th power of the local Boltzmann weight
over the infinitely deep biased Brownian valleys (\ref{vpm}).
Generalizing the approach of \cite{us_golosov}
to the biased case, we have
\begin{eqnarray} 
 (Y_k)_{valley}     
&& \sum_{\epsilon=\pm} \int_{0}^{+\infty}  dy   \left< \left( 
\frac{e^{- \beta V_{\epsilon}(\vert y \vert)} }
{ \int_0^{+\infty} dx e^{- \beta V_+(x)} 
+ \int_0^{+\infty} dx e^{-\beta V_-(x)} } \right)^k \right>_{\{V_+,V_-\}} \\
&& =
\sum_{\epsilon=\pm}     \frac{1}{\Gamma(k)}
\int_0^{+\infty} dq q^{k-1}  \left<    e^{-q \int_0^{+\infty} dx e^{- \beta V_{-\epsilon}(x)} } \right>   
   \left< \int_{0}^{+\infty}  dy  e^{- k \beta V_{\epsilon}(y)} e^{-q \int_0^{+\infty} dx e^{- \beta V_{\epsilon}(x)} } \right> 
\label{ykt}
\end{eqnarray}
Using the results (\ref{pvpinter}) of the Appendix, we finally get
\begin{eqnarray} 
&& (Y_k)_{valley}  = 
\frac{2}{\Gamma(k) \Gamma^2(1+\mu)} \left( \frac{\sigma \beta^2}{4}\right)^{k-1}
\int_0^{+\infty} ds 
\frac{ \left(\frac{s}{2} \right)^{2 \mu-1} }{  I_{\mu}^2(s) }
\int_0^s dz z^{2k-1} I_{\mu}(z)
\left[
K_{\mu}(z) 
- \frac{K_{\mu}(s)}
{ I_{\mu}(s)}
I_{\mu}(z)  \right]
\label{ykvalley}
\end{eqnarray}

\subsection{Correlation function inside a renormalized valley}
 
The correlation function of two particles 
at Boltzmann equilibrium in
an infinitely deep biased Brownian valley
reads
\begin{eqnarray} 
C_{valley}(l>0) && =   2 \sum_{\epsilon=\pm}  \int_{0}^{\infty} dy 
  \left< 
\frac{e^{- \beta V_{\epsilon}( y ) - \beta V_{\epsilon} ( y+l )} }
{ \left( \int_0^{\infty} dx e^{- \beta V_+(x)} 
+ \int_0^{\infty} dx e^{-\beta V_-(x)} \right)^2 } 
 \right>  
\nonumber \\ && +  2 \int_{0}^{l} dy 
 \left< 
\frac{e^{- \beta V_+( y ) - \beta V_-( l-y )} }
{ \left( \int_0^{+\infty} dx e^{- \beta V_+(x)} 
+ \int_0^{+\infty} dx e^{-\beta V_-(x)} \right)^2 } 
 \right>  
\label{corre2} 
\end{eqnarray}
 where the average $<..>$ is over the realizations $(V_+,V_-)$ satisfying (\ref{vpm}).

Using the explicit results of Appendix \ref{pathbias},
we finally get
\begin{eqnarray} 
&& {\hat C}_{valley} (p) 
= \frac{8 }{\Gamma^2(1+\mu) }
\int_{0}^{\infty} \frac{ds}{s} \ 
\frac{ \left( \frac{s}{2} \right)^{2\mu} }{ I_{\mu}^2 ( s )} 
   \int_0^{s} dz_1 z_1  I_{\mu} ( z_1) 
 \int_0^{s} dz_2 z_2    
  \left(   K_{\mu}(z_2 )  - \frac{ K_{\mu}(s) }
  { I_{\mu}(s) }  I_{\mu}(z_2 )   \right) \nonumber  \\
 && [ \theta(z_2-z_1)   I_{\nu} (z_1) 
 \left(   K_{\nu}(z_2 )  - \frac{ K_{\nu}(s) }{I_{\nu} (s) } I_{\nu}(z_2 )
    \right) + \theta(z_1-z_2)   I_{\nu} (z_2) 
 \left(   K_{\nu}(z_1 )  - \frac{ K_{\nu}(s) }{I_{\nu} (s) } I_{\nu}(z_1 )  
  \right) ]  
\nonumber \\ && +  \frac{4 }{\Gamma^2(1+\mu) } \int_{0}^{\infty}  \frac{ds}{s} 
 \ 
\frac{ \left( \frac{s}{2} \right)^{2\mu} }{ I_{\nu}^2 ( s )}
  \left[ \int_0^{s} dz z    I_{\nu} (z)  
 \left(   K_{\mu}(z )  - \frac{ K_{\mu}(s) } { I_{\mu}(s) } 
 I_{\mu}(z )   \right) 
   \right]^2
\label{correvalley}
\end{eqnarray}

\section{Discussion of the universality} 

\label{universal}

We now briefly discuss the question of the universality.
The RSRG method 
that describes the large scale structure at scale $\Gamma$ of the 
random potential 
is valid for all discrete models with random forces \cite{us_sinai}.
The parameter $2\delta$ that describes the distribution
of barriers against the drift at large scale (\ref{rgdrift})
may be expressed for a discrete random force model
as the non-zero solution of the equation \cite{us_sinai}
\begin{eqnarray}
\overline{e^{-2 \delta f} }=1
\end{eqnarray}
which is known to determine the anomalous diffusion exponent  $\mu=2 \delta T$
\cite{kestenetal,derrida}.
So for a given value of the parameters $(2\delta,\sigma)$,
the renormalized landscape at scale $\Gamma$ is universal.

However, it is clear from the analysis of the escape time
of a renormalized valley (\ref{theta1V})
that the prefactor (\ref{tauoV})
in front of the Arrhenius factor $e^{\beta \Gamma}$ 
is not universal  :
the partition functions $Z_V$ and $Z_B$ depend on the details
over finite scales of the potential near a bottom of a renormalized valley
and near a top of a barrier.  
 
So for a potential that belongs to the universality class $(2\delta,\sigma)$,
but that is not a biased Brownian at small scales,
the distribution of the trapping times in the renormalized landscape
at scale $\Gamma$ reads
 \begin{eqnarray}
P_{\Gamma}(\tau) && = \frac{\mu}{\tau} 
\left( \frac{ \beta e^{\beta \Gamma} }{ \tau} \right)^{\mu}
( \overline{Z_V^{\mu}} ) ( \overline{Z_B^{\mu}} )
\label{ptaugene}
\end{eqnarray}
so that the quantitative mapping onto the trap model (\ref{qttau})
is realized for the choice of the RG scale $\Gamma$ as
a function of $t$ according to
\begin{eqnarray}
\Gamma(t) = T \ln \left[ \frac{t}{\beta (\overline{Z_V^{\mu}} )^{\frac{1}{\mu}} (\overline{Z_B^{\mu}} )^{\frac{1}{\mu}} }  \right]
\end{eqnarray}
which corresponds to the length scale
\begin{eqnarray}
b(t)= b_{\Gamma(t)}  
= \frac{ 1 }{\sigma \beta^2 \mu^2} 
\  \frac{t^{\mu}}{\beta^{\mu} ( \overline{Z_V^{\mu}} ) (\overline{Z_B^{\mu} })}   \end{eqnarray}
This shows that the factor $\mu^2$ is universal and comes
from the mean length of descending bonds in the renormalized
landscape at large scale, whereas the factor $\Gamma^2(1+\mu)$
of the biased Brownian motion is not universal
and comes from the probability distribution
of the partition function of a biased Brownian valley (\ref{momentzbis}).
However it is expected to be valid for discrete models
in the limit where the lattice constant is very small as compared
to the thermal length $l_T=T^2/\sigma$.
For the localization parameters and the correlation function of two 
particles inside the same renormalized valley, the discussion of
the universality is the same as in the symmetric case \cite{us_golosov}.

\section{Conclusions and perspectives}

\label{conclusion}

To study the anomalous diffusion phase $x \sim t^{\mu}$
of the directed trap model and of the Sinai diffusion with bias,
we have extended the usual RSRG method that assumes a full localization
in a single valley
to allow for the spreading of the thermal packet over many renormalized valleys.
We have shown how all observables can be computed via a series expansion
in $\mu$ : at any given order $\mu^n$, it is sufficient 
to consider the spreading over at most $(1+n)$ traps. 
We have given explicit rules for the statistical properties of these traps.
We have shown the exactness of these expansions in $\mu$
by comparing up to order $n=2$ with the already known exact results,
such as the diffusion front \cite{annphys}, the thermal width \cite{aslangul}
and the localization parameter $Y_2$ \cite{comptejpb}. 
Our construction moreover gives a clear physical picture
of the localizations properties in the anomalous diffusion phase,
and explains the typical shape of the diffusion
front in a given sample obtained by numerical simulation
(Figure 4 of \cite{comptejpb}).

In a forthcoming paper \cite{papertrap}, we will adapt
our method to study the localization properties and the aging behaviors
in the symmetric (i.e. undirected) trap model which has attracted a lot of interest recently \cite{isopi,bertin,benarous}.

For the field of biased diffusion in one-dimensional random potentials,
it would be very interesting to study the influence
of correlations on the localization properties studied
here for the Brownian case. In particular, the case of 
 algebraic correlations 
$\overline { (U(x)-U(y))^2 } \sim \vert x-y \vert^{\gamma} $ 
is known to give rise to a creep motion for $0<\gamma<1$ \cite{correalge}.
For DNA sequences, it seems that the interesting cases
are not only the Brownian case $\gamma=1$ \cite{lubensky}
but also the values $\gamma >1$ \cite{arneodo}. 
Another physically interesting case concerns
the logarithmic correlations, which give rise to a freezing transition
in the dynamics \cite{castillo} 
as well as in the statics
\cite{carpentier}.

From the point of view of the RSRG method,
 since the usual RSRG is asymptotically exact for
infinite-disorder fixed points \cite{daniel}, 
the extension introduced here can be seen as a systematic expansion
in the inverse disorder strength. It can therefore be used in
the field of random quantum spin chains \cite{daniel}
to study the Griffith phases, 
and for the classical random field Ising chain
in the presence of a small magnetic external field \cite{us_rfim}.

Finally, the expansion in the important traps
for the dynamical models 
discussed in this paper has a static counterpart, 
with the following differences :
in the static case, the expansion parameter is the temperature $T$,
and the main trap $M$ corresponds to the absolute minimum of the random potential. We have already shown in \cite{us_toy} 
for the toy model consisting of a Brownian potential plus a quadratic potential,
how the thermal cumulants at first order in $T$ can be explained
by studying the statistical properties of the configurations
presenting two nearly degenerate minima.  
We will discuss in \cite{us_identities} in a more general context
the structure of the low-temperature series expansions in some disordered
systems.

\begin{acknowledgments}

 I thank Pierre Le Doussal for helpful discussions at the early stage of
this work and for his remarks on the manuscript.

\end{acknowledgments}

\appendix

\section{ Useful properties of the L\'evy diffusion front for $0<\mu<1$ }

\label{levy}

In this appendix, we recall some useful properties of the 
L\'evy distributions (\cite{annphys,montroll} and references therein). 

\subsection{Definition and properties of one-sided Levy stable laws }

The rescaled sum 
\begin{eqnarray}
y = \frac{1}{n^{\frac{1}{\mu} } } \sum_{i=1}^n t_i
\end{eqnarray}
of $n$ identical independent positive random variables distributed
with a law presenting the algebraic decay
\begin{eqnarray}
p(t) \opsimeq_{t \to \infty} \frac{A}{t^{1+\mu} }
\label{alge}
\end{eqnarray}
where $0<\mu<1$, has for limit distribution
as $n \to \infty$ the one-sided Levy law $L_{\mu,c(\mu;A)}(y)$
defined by its Laplace transform
\begin{eqnarray}
\int_0^{+\infty} dy e^{-s y} L_{\mu,c}(y) = e^{- c s^{\mu}} 
\label{laplacelevy}
\end{eqnarray}
and where the constant $c$ reads
\begin{eqnarray}
c(\mu;A) = \frac{ \pi A}{ \sin \pi \mu \Gamma(1+\mu) }
\label{cmuA}
\end{eqnarray}
In this paper, we will only use
the following series representation \cite{annphys,montroll} 
\begin{eqnarray}
 L_{\mu,c}(y) = - \frac{1}{\pi y} \sum_{k=1}^{+\infty} 
\left( -\frac{c}{y^{\mu}} \right)^k 
\frac{\Gamma(1+k \mu)}{\Gamma(1+k)} 
\sin \pi \mu k
\label{serielevy}
\end{eqnarray}
which is convergent in the whole phase $0<\mu<1$.

We stress here that we have defined the constant $c$ 
by the Laplace transform (\ref{laplacelevy}). Writing the inverse Laplace transform as a Fourier integral yields
\begin{eqnarray}
L_{\mu,c}(y)  = \int_{-i \infty}^{+i \infty} \frac{ds}{2i \pi} e^{-s y- c s^{\mu}} = \int_{- \infty}^{+ \infty} \frac{dt}{2 \pi} e^{-i t y
- c t^{\mu}(\cos \frac{\pi \mu}{2} +i {\rm sgn} \sin \frac{\pi \mu}{2} ) }  \end{eqnarray}
so that the constant ${\cal C}$ appearing in the usual Fourier transform
of Levy distributions
\begin{eqnarray}
L_{\mu,c}(y) && = \int_{- \infty}^{+ \infty} \frac{dt}{2 \pi} e^{-i t y
- {\cal C} t^{\mu}(1 +i {\rm sgn} (t) \tan \frac{\pi \mu}{2} ) }  
\label{fourierlevy}
\end{eqnarray}
reads in terms of the Laplace constant $c$
\begin{eqnarray}
 {\cal C} = c  \cos \frac{\pi \mu}{2} 
\label{cfourier}
\end{eqnarray}

\subsection{ L\'evy diffusion front for the trap model}

For a given trap $\tau$, the distribution of the escape time $t$ 
is exponential
\begin{eqnarray}
f_{\tau}(t) = \frac{1}{\tau} e^{- \frac{t}{\tau} }  
\end{eqnarray}
which yields after averaging over $\tau$  (\ref{lawrg})
\begin{eqnarray}
\overline{ f_{\tau}}(t)  = \int_0^{+\infty} d\tau q(\tau) f_{\tau}(t)
= \int_0^{+\infty} \frac{dv}{v} q \left( \frac{t}{v} \right) e^{-v} 
\opsimeq_{t \to \infty}  \frac{ \mu \Gamma(1+\mu)}{t^{1+\mu}}
\label{meanf}
\end{eqnarray}

For a given sample $(\tau_0,\tau_1,...)$, the probability $P_t(n)$ 
for the particle to be in the trap $n$ at time $t$ reads
\begin{eqnarray}
P_t(n) = \int \prod_{i=0}^{+\infty} dt_i f_{\tau_i} (t_i) \theta(t_0+t_1...+t_{n-1}<t< t_0+t_1+...+t_{n} ) 
\end{eqnarray}
The average over the disorder 
\begin{eqnarray}
\overline{P_t(n)} = \int \prod_{i=0}^{+\infty} dt_i
 \overline{f_{\tau} }(t_i) \theta(t_1+t_2+...+t_{n-1}<t< t_1+t_2+...+t_{n} ) 
\end{eqnarray}
shows that the diffusion front at large time is 
directly related to the properties of the sum of a large number $n$
of independent variables $t_i$ distributed with
the law (\ref{meanf}) presenting an algebraic decay (\ref{alge}) :
 the rescaled variable $y=\frac{t}{n^{1/\mu}}$
is distributed with a 
one-sided stable Levy distribution $L_{\mu,c_{trap}(\mu)}$
(\ref{laplacelevy}),
 where the constant $c_{trap}(\mu)$ reads for the case (\ref{meanf},\ref{cmuA})
\begin{eqnarray}
 c_{trap}(\mu) = \frac{\pi \mu} { \sin \pi \mu }   
\label{ctrap}
\end{eqnarray}
The variable $X=\frac{n}{t^{\mu}}= y^{-\mu}$
is thus distributed with the law
\begin{eqnarray}
f_{\mu,c}(X)=\frac{1}{\mu X^{1+\frac{1}{\mu}} } L_{\mu,c}(X^{-\frac{1}{\mu}})
\label{front}
\end{eqnarray}
with the special value $c=c_{trap}(\mu)$.

In particular, the series expansion
(\ref{serielevy}) 
gives the following series representation for the diffusion front
\begin{eqnarray}
f_{\mu,c}(X)  =  \frac{c}{ \pi \mu } 
 \sum_{k=1}^{+\infty} 
\left( - c X \right)^{k-1} \frac{\Gamma(1+k \mu)}{\Gamma(1+k)} 
\sin \pi \mu k   = c 
 \sum_{k=1}^{+\infty} 
 \frac{ \left( - c X \right)^{k-1} }{ (k-1)! \Gamma(1-k \mu)}  
\label{fserie}
\end{eqnarray}

Using the series expansion
\begin{eqnarray}
\frac{1}{ \Gamma(1-z)} = 1+\sum_{m=1}^{+\infty} d_m 
(-1)^m z^m 
\end{eqnarray}
with the first coefficient
\begin{eqnarray}
d_1 && =\gamma_{E} \\
d_2 && = \frac{\gamma_{E}^2}{2}- \frac{\pi^2}{12} \\
\label{dvalues}
\end{eqnarray}
where $\gamma_E$ denotes the Euler's constant, we get the
expansion in $\mu$ of the series (\ref{fserie}) for fixed $c$ 
\begin{eqnarray}
f_{\mu,c}(X)  = c e^{-c X }\left[ (1-d_1 \mu + d_2 \mu^2 )
+ ( d_1 \mu -3 d_2 \mu^2 ) c X  
+ d_2 \mu^2 ( c X )^2  +O(\mu^3) \right] 
\label{seriefmu}
\end{eqnarray}

Expanding also in $\mu$ the value (\ref{ctrap})
\begin{eqnarray}
 c_{trap}(\mu)  
= 1 + \frac{\pi^2}{6} \mu^2 + O(\mu^3)
\end{eqnarray}
we get that the diffusion front reads up to second order in $\mu$ 
\begin{eqnarray}
g(X) = f_{\mu,c_{trap}(\mu)}(X) 
  =  [1- d_1 \mu + (d_2+ \frac{\pi^2}{6}) \mu^2 ] e^{-  X  }
+ [ d_1 \mu - (3 d_2 + \frac{\pi^2}{6} ) \mu^2 ]  X  e^{- X } 
+ d_2 \mu^2   X^2 e^{- X } +O(\mu^3) 
  \label{ffmu}
\end{eqnarray}
Using the numerical values (\ref{dvalues}), we finally get
the expression (\ref{fmu}) of the text.  

\subsection{L\'evy diffusion front for the biased Sinai model}

The exact form of the diffusion front was first determined in
\cite{kestenetal} for a corresponding discrete model.
For the continuum model, the result has been proved in
the theorem 1 of \cite{tanaka}, which states that,
for $0<\mu<1$, the rescaled variable $X=\frac{x}{t^{\mu}}$ has for probability distribution (\ref{front}),
 where the constant $c_s(\mu)$ is given by a complicated
implicit expression in \cite{tanaka}.
The value of this constant has been proven 
in \cite{hushiyor} to have the following simple expression
\begin{eqnarray}
 c_{sinai}(\mu) = 8^{\mu}  \frac{ \pi \mu}{ 2 \Gamma^2(\mu) \sin \pi \mu }
\label{explicmu}
\end{eqnarray}
where we have used (\ref{cfourier}).

In \cite{annphys}, the same form was conjectured from the 
heuristic equivalence 
with the directed trap model via an indentification
of the parameters on some observable
\begin{eqnarray}
 c_{sinai}(\mu) 
= \frac{2}{x_1} \left( \frac{ \tau_1   }{ 2} \right)^{\mu}
 \frac{ \pi \mu}{   \Gamma^2(\mu) \sin \pi \mu }
\label{explicmuann}
\end{eqnarray}
where $x_1=2 T^2/\sigma$  
and $\tau_1=x_1^2/(2 D_0)$
in terms of the diffusion constant $D_0$ in the pure case.
This expression indeed coincides with (\ref{explicmu}) for the units 
$T=1$, $\sigma=\frac{1}{2}$ and $D_0=\frac{1}{2}$ used in \cite{hushiyor}.

In the notations used in this article $D_0=T$ (\ref{fokkerplanck}),
this corresponds to
\begin{eqnarray}
 c_{sinai}(\mu) 
=   \frac{ \sigma \beta^2   }{ (\sigma^2 \beta^3)^{\mu} } 
 \frac{ \pi \mu}{   \Gamma^2(\mu) \sin \pi \mu }
\label{explicrsrg}
\end{eqnarray}

To compare with the directed trap model, it is convenient to consider the ratio
of the two constants (\ref{ctrap})
\begin{eqnarray}
\frac{ c_{sinai}(\mu) }{c_{trap}(\mu) } 
=  \frac{ \sigma \beta^2   }{ (\sigma^2 \beta^3)^{\mu} } 
 \frac{ 1}{   \Gamma^2(\mu)  }
\label{ratio}
\end{eqnarray}
So beyond the natural dimensional factors, there is still a function $\Gamma^2(\mu)$
between the two models, whose origin will be discussed in details in
the text.

\section{ Statistics of the internal structure of renormalized valleys }

\label{pathbias}

\subsection{Distribution inside a renormalized valley}

To compute the functionals (\ref{boltzmannvalleybias}),
we generalize the approach developed in \cite{us_golosov}
for the symmetric case $\mu=0$.
We first exponentiate the denominator  
\begin{eqnarray}
P_{\infty}(y>0) && = \int_0^{\infty} dq R_{\infty}^-(q) S_{\infty}^+(y,q)  \\
P_{\infty}(y<0) && = \int_0^{\infty} dq R_{\infty}^+(q) S_{\infty}^-(y,q)
\label{pytbias}
\end{eqnarray}
where
\begin{eqnarray}
R_{\Gamma}^{\pm}(q) \equiv \left<    e^{-q \int_0^{l_{\Gamma}} dx e^{- \beta V_{\pm}(x)} } \right>_{\{V_{\pm}\}}
\label{rqdef}
\end{eqnarray}
\begin{eqnarray}
S_{\Gamma}^{\pm}(y,q) && \equiv \left<     e^{- \beta V_{\pm}(y)} e^{-q \int_0^{l_{\Gamma}} dx e^{- \beta V_{\pm}(x)} } \right>_{\{V_{\pm}\}} 
\label{syqdef}
\end{eqnarray}

These two functionals may be expressed as 
\begin{eqnarray}
R_{\Gamma}^{\pm}(q)  = {\cal N_{\pm}} \int_0^{+\infty} dl \lim_{\epsilon \to 0} \frac{1}{\epsilon^2} F_{[0,\Gamma]}^{\pm}(\Gamma-\epsilon,l \vert \epsilon)
\label{rqtbias}
\end{eqnarray}
\begin{eqnarray}
S_{\Gamma}^{\pm}(y,q)  = {\cal N_{\pm}} \int_0^{\Gamma} du \int_0^{+\infty} dl \lim_{\epsilon \to 0} \frac{1}{\epsilon^2} 
 F_{[0,\Gamma]}^{\pm}(\Gamma-\epsilon,l|u)  e^{-\beta u} F_{[0,\Gamma]}^{\pm}(u,y|\epsilon)
\label{syqtbias}
\end{eqnarray}
in terms of the path-integrals
\begin{eqnarray}
F_{[0,\Gamma]}^{\pm}( u , l \vert u_0)
&& = \int_{V(0)=u_0}^{V(l)=u}
{\cal D }V(x) e^{-\frac{1}{4 \sigma } \int_0^{l} dx 
\left( \frac{d V} {dx} \pm F_0  \right)^2 -q \int_0^{l} dx e^{- \beta V(x)}}
\Theta_{[0,\Gamma]} \{V(x)\}
\end{eqnarray}
where the symbol $\Theta_{[0,\Gamma]} \{V(x)\}$ means that there are absorbing boundaries at $V=0$ and $V=\Gamma$.
The expansion of the quadratic term of the measure yields
\begin{eqnarray}
F_{[0,\Gamma]}^{\pm}( u , l \vert u_0)
  = e^{- \frac{F_0^2}{4 \sigma} l \mp  \delta (u-u_0) }
F_{[0,\Gamma]}( u , l \vert u_0)
\end{eqnarray}
where 
\begin{eqnarray}
F_{[0,\Gamma]}( u , l \vert u_0)  = 
\int_{V(0)=u_0}^{V(l)=u}
{\cal D }V(x) e^{-\frac{1}{4 \sigma} \int_0^{l} dx 
\left( \frac{d V} {dx} \right)^2 -q \int_0^{l} dx e^{- \beta V(x)}}
\Theta_{[0,\Gamma]} \{V(x)\} \\
\end{eqnarray}
represents the analogous path-integral for the symmetric case.
Its Laplace transform has been computed in Equation (B18) of \cite{us_golosov}).
We get 
\begin{eqnarray}
R_{\Gamma}^{\pm}(q) && = \frac{ {\cal N_{\pm}} e^{\mp \delta \Gamma} }
  {\sigma E(0,\Gamma, \frac{F_0^2}{4 \sigma})}  
\label{resrqtbias}
\end{eqnarray}
and the Laplace transforms with respect to $y$
\begin{eqnarray}
{\hat S}_{\Gamma}^{\pm}(p,q)  \equiv \int_0^{+\infty} dy e^{-py}
S_{\Gamma}^{\pm}(y,q)  = \frac{\cal N_{\pm} e^{\mp \delta \Gamma} }{\sigma^2}
 \int_0^{\Gamma} du   e^{-\beta u}   \frac{E(0,u,\frac{F_0^2}{4 \sigma}) }
{E(0,\Gamma,\frac{F_0^2}{4 \sigma})}
 \frac{ E(u,\Gamma,p+\frac{F_0^2}{4 \sigma})}{E(0,\Gamma,p+\frac{F_0^2}{4 \sigma})} 
\label{spqtbias}
\end{eqnarray}
in terms of the function 
\begin{eqnarray}
E(u,v,p) = \frac{2}{\beta} \left[ I_{\frac{2}{\beta}  \sqrt{\frac{p}{\sigma}}}
\left(\frac{2}{\beta}  \sqrt{\frac{q}{\sigma}} e^{- \frac{ \beta u}{2}} \right)
 K_{\frac{2}{\beta}  \sqrt{\frac{p}{\sigma}}}
\left(\frac{2}{\beta}  \sqrt{\frac{q}{\sigma}} e^{- \frac{ \beta v}{2}} \right) 
- K_{\frac{2}{\beta}  \sqrt{\frac{p}{\sigma}}}
\left(\frac{2}{\beta}  \sqrt{\frac{q}{\sigma}} e^{- \frac{ \beta u}{2}} \right) 
I_{\frac{2}{\beta}  \sqrt{\frac{p}{\sigma}}}
\left(\frac{2}{\beta}  \sqrt{\frac{q}{\sigma}} e^{- \frac{ \beta v}{2}} \right)\right]
\label{defE}
\end{eqnarray}

The normalizations ${\cal N_{\pm}}$ are obtained with
the conditions $R_{\Gamma}^{\pm}(q \to 0)=1$   
\begin{eqnarray}
{\cal N_{\pm}} = \sigma \frac{ \sinh \delta \Gamma}{ \delta}  e^{\pm \delta \Gamma}
\label{normapm}
\end{eqnarray}
We thus obtain that there is no dependence in the sign $\pm$
for the functionals $R_{\Gamma}^{\pm}(q)$ and ${\hat S}_{\Gamma}^{\pm}(p,q)$.
As a consequence we get the non-intuitive result that the probability
distribution $P_V(y)$ is symmetric in $y \to -y$.
The restoration of this symmetry comes from the conditioning
of the random walk to reach $\Gamma$.
We note that similarly, the distribution of the random times $l_{\Gamma}^{+}$
and $l_{\Gamma}^{-}$ are also the same, since we have,
with the notations of \cite{us_sinai}
\begin{eqnarray}
\frac{P_{\Gamma}^{\pm}(\zeta=0,s)} {P_{\Gamma}^{\pm}(\zeta=0,0)}
= \frac{U_{\Gamma}^{\pm}(s)} {U_{\Gamma}^{\pm}(0)}
= \frac{ { \sqrt {s+\delta^2} } \sinh \delta\Gamma  }
{ \delta \sinh \Gamma \sqrt {s+\delta^2} }
\end{eqnarray}

The Laplace transform of the distribution inside a valley thus reads
\begin{eqnarray}
{\hat P}_V(p) && \equiv \int_0^{+\infty} dy e^{-p y} P_{V}(y>0)  =
 \int_0^{\infty} dq R_{\infty}^{\pm}(q) S_{\infty}^{\pm}(p,q) \\
 R_{\infty}^{\pm}(q) && =\frac{1}{\Gamma(1+\mu ) }
\frac{ \left( \frac{1}{\beta} \sqrt{\frac{q}{\sigma}} \right)^{\mu}}
{ I_{\mu}\left(\frac{2}{\beta} \sqrt{ \frac{q}{\sigma} } \right)} \\
{\hat S}_{\infty}^{\pm}(p,q) && = 
\frac{\left( \frac{1}{\beta} \sqrt{\frac{q}{\sigma}}
\right)^{\mu} }{ q \Gamma(1+\mu ) }
\int_0^{\frac{2}{\beta}  \sqrt{\frac{q}{\sigma} }} dz z 
   \frac{ I_{\sqrt{\mu^2+ \frac{4 T^2 p}{\sigma}}} (z) }
{I_{\sqrt{\mu^2+ \frac{4 T^2 p}{\sigma}}} 
\left(\frac{2}{\beta} \sqrt{ \frac{q}{\sigma} }  \right)} 
\left[K_{\mu}(z) 
- \frac{K_{\mu} \left(\frac{2}{\beta} \sqrt{ \frac{q}{\sigma} } \right)}
{ I_{\mu} \left(\frac{2}{\beta} \sqrt{ \frac{q}{\sigma} } \right)} I_{\mu}(z)  \right]
\label{pvpinter}
\end{eqnarray}

The final result is thus given by Equation (\ref{pvpfinal}) of the text.

\subsection{Partition function of a renormalized valley}

We now consider the probability distribution of the partition function
of a renormalized valley (\ref{zvdef})
\begin{eqnarray}
Z_V = \int_0^{+\infty} dz e^{-\beta V_-(z)}
+ \int_0^{+\infty} dz e^{-\beta V_+(z)} 
\end{eqnarray}
where the potentials satisfy the constraints (\ref{vpm}).

Its Laplace transform can be directly expressed
in terms of the functions (\ref{rqdef})
which we have computed before (\ref{pvpinter})
\begin{eqnarray}
\int_0^{+\infty} dZ_V {\cal P} (Z_V) e^{-q Z_V}  = R_{\infty}^{+}(q) R_{\infty}^{-}(q)  = \left[ \frac{1}{\Gamma(1+\mu ) }
\frac{ \left( \frac{1}{\beta} \sqrt{\frac{q}{\sigma}} \right)^{\mu}}
{ I_{\mu}\left(\frac{2}{\beta} \sqrt{ \frac{q}{\sigma} } \right)} \right]^2
\label{emqzv}
\end{eqnarray}
After the rescaling $Z_V=z_1/(\sigma \beta^2)$, this corresponds to the 
result (\ref{laplacez}) given in the text.

\end{document}